\documentclass[unnumsec,webpdf,modern,large]{oup-authoring-template}

\newcommand{\REV}{}

\graphicspath{{Fig/}}

\theoremstyle{thmstyleone}%
\theoremstyle{thmstyletwo}%
\theoremstyle{thmstylethree}%

\newcommand\invisiblesection[1]{%
  \refstepcounter{section}%
  \addcontentsline{toc}{section}{\protect\numberline{\thesection}#1}%
  \sectionmark{#1}}

\begin{document}


\journaltitle{Journal of the Royal Statistical Society, Series A (Statistics in Society)}
\DOI{DOI to follow}
\copyrightyear{2024}
\pubyear{2024}
\access{}
\appnotes{Preprint article}

\firstpage{1}


\title[Modelling first arrivals of migratory birds with citizen science data]{Extreme-value modelling of migratory bird arrival dates: Insights from citizen science data}

\author[1,$\ast$]{Jonathan Koh\ORCID{0000-0002-7287-0588}}
\author[2]{Thomas Opitz\ORCID{0000-0002-5863-5020}}

\authormark{Koh and Opitz}
\address[1]{\orgdiv{Institute of Mathematical Statistics and Actuarial Science}, \orgname{Oeschger Centre for Climate Change Research, University of Bern}, 
\country{Switzerland}
}
\address[2]{\orgdiv{Biostatistics and Spatial Processes} (UR546), \orgname{INRAE}, Avignon, 
\country{France}}

\corresp[$\ast$]{Corresponding author. \href{jonathan.koh@unibe.ch}{jonathan.koh@unibe.ch}}

\received{Date}{0}{Year}
\revised{Date}{0}{Year}
\accepted{Date}{0}{Year}



\abstract{
Citizen science mobilises many observers and gathers huge datasets but often without strict sampling protocols, resulting in observation biases due to heterogeneous sampling effort, which can lead to biased statistical inferences. We develop a spatiotemporal Bayesian hierarchical model for bias-corrected estimation of  arrival dates of the first migratory bird individuals at a breeding site. Higher sampling effort could be correlated with earlier observed dates. We implement data fusion of two citizen-science datasets with fundamentally different protocols (BBS, eBird) and map posterior distributions of the latent process, which contains four spatial components with Gaussian process priors: species niche; sampling effort; position and scale parameters of annual first arrival date. The data layer includes four response variables: counts of observed eBird locations (Poisson); presence-absence at observed eBird locations (Binomial); BBS occurrence counts (Poisson);  first arrival dates (Generalised Extreme-Value). We devise a Markov Chain Monte Carlo scheme and check by simulation that the latent process components are identifiable. We apply our model to several migratory bird species in the northeastern US for 2001--2021 and find that the sampling effort significantly modulates the observed first arrival date. We exploit this relationship to effectively bias-correct predictions of the true first arrivals.
}



\maketitle

\section{Introduction}

\subsection{The rise of citizen science}

Defining citizen science and its boundaries is difficult \citep{Haklay.2021}. Broadly, the field involves a collaborative approach to scientific inquiry that engages volunteers and non-professionals in the collection, analysis, and interpretation of data. This participatory model is not new. Friedrich Wilhelm Herschel, a musician by training, discovered the planet Uranus in 1781 using his telescopes. Charles Darwin conducted crowd-sourcing projects in the $19^\text{th}$ century by recruiting acquaintances and travellers to write to him about their observations \citep{Browne1996.darwin}. The longest-running citizen science initiative is the annual Christmas Bird Count \citep{Bock.1981}, a census administered by the National Audubon Society of wintering birds in the Western Hemisphere by volunteer birdwatchers since 1900.

Citizen science has gained significant traction over the past two decades, spurred on by growing world literacy levels and the advent of devices and infrastructures in information technology for gathering, reporting, sharing and storing data \citep{Newman.2012, Wynn_book.2017}. These datasets are often collected at relatively low cost and sometimes with unconventional funding sources \citep{SILVERTOWN.2009}. Zooniverse\footnote{\url{https://www.zooniverse.org/}}, the world's largest platform for volunteer-based research, has seen a surge in registered volunteers to over 1.9 Million since its inception in 2009. 
Climate\textit{prediction}.net runs climate modelling experiments using the home computers of thousands of volunteers. There is now a peer-reviewed journal \citep{Bonney-2016} dedicated to disseminating research on  citizen science, and the 2019 Citizen Science Association biannual conference attracted 818 registered delegates from 28 countries. 

Citizen science is expected to grow in importance. Technological progress will improve how machines and citizen science work together; for example, volunteer-classified training sets have already been used to improve the performance of machine learning approaches in astronomy \citep{Marshall.2015}. \citet{Frisl.2020.UN} note that traditional data sources are not sufficient for measuring the United Nations Sustainable Development Goals \citep{UN.2015}, but sources from citizen science are required to better inform policies and actions. Citizen-science initiatives can also have community-level impacts on the participants \citep{Jordan.2012}, such as by empowering them and giving them a voice in local environmental decision-making \citep{Bonney2016CanCS} while increasing public awareness of the scientific process. Though citizen science's global economic value has been estimated to amount to over  2.5 billion USD annually, the majority of data collected through citizen-science initiatives do not reach analysis in peer-reviewed literature \citep{THEOBALD.2015}. There is a need to develop and disseminate statistical methods that facilitate wider scientific use of these datasets. 


\subsection{Analysis of data for species monitoring}

Much of the recent growth in citizen science has occurred in the ecological and environmental sciences \citep{McKinley.2015, Pocock.2018, Fraisl.2022}. A prominent example in ecology is the eBird project \citep{eBird}, launched in 2002, which has led to a database of over 200 Gigabytes providing information on over 600 million bird observations. Another important bird monitoring program within the scope of citizen science is the North American Breeding Bird Survey (BBS). It was launched in 1966 for official monitoring purposes and engages large numbers of trained volunteers in collecting standardised data on bird populations  according to a relatively strict sampling protocol along several selected routes and during each year. BBS data can be considered of high quality but sampling is relatively sparse in space and occurs only during fixed periods in the year, whereas the sampling of eBird data is much more heterogeneous but provides a generally denser spatio-temporal coverage, as we will detail in Section~\ref{sec:data}. We exploit these two data sources in this work. Regarding non-avian species, another recent example among many others is the deployment of camera-trapping projects for mammal monitoring \citep{Hsing.eMammal.2022}.

Many observation data in ecology are \emph{opportunistic}, i.e., they were collected incidentally or without a predefined research question in mind, so information on the criteria applied by the observer for sampling and reporting observations is limited. However, opportunistic data often allow for a spatiotemporal coverage of species monitoring that would not be attainable with protocol-based data collected by professionals at relatively sparse and deterministically chosen sampling locations. Citizen-science data is particularly valuable for statistical inference on low-probability events (e.g., occurrences of rare species, or unusual or extreme phenological events) and on occurrence times of events, such as the arrival of migratory birds at their breeding site in spring.  
The general data quality from citizen-science initiatives is high \citep{Kosmala.2016}, though the need for statistical methods to account for different data biases prevails. \citet{Isaac.2014} discuss bias-correction approaches for ecological trend estimates, and conclude that opportunistic data would be further enhanced if information on the sampling effort at the data collection points can be captured. They further identify four key aspects of biases induced by volunteer sampling, due to uneven sampling in time, space, effort per visit, and uneven detectability of species individuals, which can vary by observer, species and land cover. For example,  detectability could be higher in open landscapes than in dense forests. Moreover, the expertise of individual observers for detecting and identifying species may increase over time due to increasing experience and knowledge-sharing with other observers, thus reducing observation biases \citep{Kelling2015,Johnston.2018}. 

Using eBird, statistical  modelling of sampling effort expressed through various criteria was discussed by \cite{tang.2021}. More generally, a large body of literature has emerged to characterise heterogeneous sampling effort and quantify it from available data \citep[e.g.,][]{Gelfand2019,Fink2020,Johnston2021,Johnston2023}. 

Another problem arises when only the presence of species is reported but not  their absence (\emph{presence-only data}), so areas and times without any reported occurrences could correspond either to the true absence of the species or to the presence of the species but an absence of observers. The BBS and eBird datasets do not explicitly report absences, but both rely on observation protocols requiring that all species detected at the sampled location and identified by the observer be reported. In eBird, data entries satisfying this protocol are known as \emph{checklist data}, with a unique space-time coordinate associated with each checklist; these account for the vast majority of eBird entries. Checklist data in eBird are quality-controlled by experts, and non-expert observers can train in birding with resources offered by eBird, such as tutorials, online courses and bird identification apps. Observer skill for detecting and correctly identifying various bird species still varies across observers, even for experts. Due to the large number of observers contributing to the eBird dataset, our model will capture an observation effort that represents the average observation skill.

We adopt the common approach of considering species not reported within a checklist as being absent from the space-time location of the checklist. By systematically adding such (pseudo-)absence information to the dataset, we obtain a binary observation for each combination of checklist and species, with $1$ corresponding to presence and $0$ corresponding to (pseudo-)absence of the species. We can then estimate the species presence probability using available predictors with a binomial likelihood.

\subsection{Extreme-value analysis for migratory bird arrivals}

Extreme-Value Theory \citep[EVT,][]{Coles.2001} is a branch of probability theory and statistics that originated in the first half of the 20th century. It provides a theoretically justified framework to model extreme events, i.e., the tails of data distributions. It has been extensively used in fields such as finance and environmental sciences, especially for climate data. Applications in ecology have so far mostly focused on extremes of abiotic environmental processes  \citep{Gaines1993,Katz2005}, which can have a strong influence on biotic processes, for example when very low winter temperatures contribute to limiting the extent of outbreaks of forest pests \citep{Thibaud2016}.  However, direct applications of EVT to biotic variables are scarce, due perhaps to the strongly discretised nature of species observation data, often collected in the form of presence-absence information or relatively small occurrence counts. By contrast, standard EVT is more focused on variables measured on a continuous scale. Recent studies leveraging EVT aim to model species accumulation curves \citep{borda-de-agua.2021}, extremes of species movements \citep{Wijeyakulasuriya2019} or the first arrival dates of migratory birds \citep{Wijeyakulasuriya.2023}. However, the aforementioned approaches do not explicitly consider the specific observation biases inherent to ecological datasets, which is the focus of our work here. 


First arrival dates of migratory birds at their destination are an example of  phenological  events, i.e., of recurring life history events of species, such as migration or breeding. Bird migration is a complex phenomenon taking place over large geographic scales \citep{Somveille2015}, with possible temporal trends in migration patterns resulting from changes in climate, land cover and other ecological and evolutionary factors \citep{Cotton.2003,Conklin.etal.2021}. The study of migration patterns, and particularly of arrival times at the breeding site, has attracted strong interest \citep{Linden2011,Youngflesh2021}. Approximately 20\% of all bird species are concerned by migration, which allows birds to adapt to seasonal cycles in climatic stresses and in the availability of resources such as food, especially during their breeding period. 

Previous studies often use rather complex approaches to define and estimate a date that can be viewed as representative of the arrival of birds at their breeding site. For example, \citet{Youngflesh2021} based the representative date for a given area and year on both the first arrival date of an individual bird  and the first local maximum of the species' detection probabilities estimated from eBird data. In our approach, we focus only on the first arrival date for each pixel of a spatial mesh covering the study area, i.e., we model the minimum of all dates of occurrence of the species during the year. We  use the Generalised Extreme-Value distribution motivated by EVT to model this sample extreme of all  observed occurrence times in the year, similar to \citet{Wijeyakulasuriya.2023}. Here we consider dates as a continuous variable, which is sensible since the distribution of observed arrival dates spans over a long enough period so that discretisation effects have a weak influence; Figure~\ref{fig:explore:preferential} suggests that observed dates span over several months. Although distributions for discrete extremes have been proposed in the literature \citep[e.g.,][]{Prieto2014,Hitz2017,Ranjbar2022}, they usually come with overhead for numerical computations and modelling, and asymptotic theory is limited to heavy-tailed variables, by contrast with our setting where tails have a natural finite bound and are therefore light-tailed.

In the northeastern US, 
birds arrive during spring from more southern regions where they have spent overwintering. Our goal is to analyse how the probability distribution of the date of the first arrivals of birds varies across species and space in this region, and in response to factors related to climate and land cover.  The mapping of the arrival at breeding sites should be restricted to locations where birds are present during the year, i.e., locations that provide a suitable environment for birds to breed, so the considered locations must be part of the spatial area forming the ecological niche occupied by the bird species for breeding. We thus design our statistical model to predict both the niche and the first arrival times. At locations where the species presence probability is very low according to the niche model, we will not map arrival times.

To our knowledge, the modelling approach developed here is the first that aims to appropriately capture the interplay of the niche, the sampling effort and the observed first arrival dates to provide bias-corrected predictions of the true first arrival dates. 
\citet{Wijeyakulasuriya.2023} \REV{used only eBird data and focused on modelling the spatial dependence among the first arrival dates for each year using the class of max-infinitely divisible models, so that first arrival dates at locations without observations during certain years can be predicted. We extend their approach and} consider the niche and the sampling effort when predicting first arrival dates, and we focus on revealing spatial patterns that remain stable across the whole study period.  The approach of \citet{Youngflesh2021} is based on a different metric for the arrival dates, with a less direct and intuitive interpretation, and it does not  account for sampling effort except for choosing a study area with a relatively high overall sampling effort in eBird.  

\subsection{Bayesian modelling of complex ecological data}

Observation data in ecology often do not directly measure the latent (i.e., not directly observed) processes of interest, such as the ecological niche, the sampling effort and the timing of a phenological event in our case. This is due to observation biases and complex interactions among such processes. Moreover, different data sources (e.g., eBird and BBS) can contribute complementary information about the same latent process (e.g., the ecological niche). The data we use here provide a concrete illustration. The eBird checklists  are available for 
spatiotemporal locations chosen by the observers and provide generally good spatiotemporal coverage  of the study area, especially for the most recent years in the study period. By contrast, the BBS observations are only available along predefined routes of length around 40~km, with up to 50 stops separated by around $800$~m at which observers can report occurrence numbers of detected bird species. Time intervals for observation are also imposed by the study protocol. Therefore, observation always takes place at prespecified spatiotemporal locations in BBS, with the land-cover type at those locations marked by the presence of a usually relatively large road. As a consequence, BBS data may be less representative of all possible land-cover types in comparison to eBird, and also provide no direct information about events taking place outside the prescribed observation interval, such as first arrivals.  On the other hand, within BBS many routes have been sampled during each year since the 1970s, so the temporal coverage of BBS over the full study period is more homogeneous and complete than eBird. In summary, combining information from both datasets offers the possibility of improved inferences for properties of the niche and of phenological events of bird species. 

With Bayesian hierarchical models \citep[BHMs,][]{Banerjee2003}, data are assumed to be generated conditional on latent processes, which in turn are conditioned on hyperparameters (e.g., variance or spatial correlation range). This makes it possible to account for spatial patterns, complex relationships and uncertainties. Using Bayes' Theorem, the combination of prior information based on expert knowledge and the data likelihood results in posterior distributions for latent processes and hyperparameters that reflect updated beliefs about the processes of interest given the observed data \REV{\citep{vandeSchoot2021}}.

An interesting application of BHMs in ecology is the fusion of presence-only data, available through large datasets but with strongly heterogeneous and unknown sampling effort, with presence-absence data, available through smaller datasets but with known sampling effort, to infer species distribution maps \citep[e.g.,][]{Gelfand2019}. An important element of such approaches is the inclusion of a latent Gaussian field that represents spatial heterogeneity in preferential sampling, similar to the foundational work of \citet{Diggle2010}. 

To infer the complex BHM we design for mapping the niche, observation effort and first arrival dates, we devise a Markov Chain Monte Carlo scheme with Vecchia likelihood approximations \REV{\citep{Vecchia.1988,Katzfuss.Guinness.2021}} and Metropolis-adjusted Langevin Algorithm \REV{\citep[MALA,][]{Roberts.Rosenthal.2001}} updates for the latent Gaussian fields. Our approach marks the first use of the Vecchia approximation in  Bayesian hierarchical modelling for spatial extremes, extending other recent approaches in EVT \citep{Huser.vecchia.2023, Majumder.2024}.

\subsection{Outline of the paper}

Section~\ref{sec:data} presents the datasets and extensive preprocessing steps. In Section~\ref{sec:model}, we recall the basics of EVT and introduce the Bayesian hierarchical model we use to identify the spatial variability of three aspects (sampling effort; niche of a given species; date of arrival of the first individuals during spring migration for a given species). Details of latent Gaussian process models and Markov-Chain Monte-Carlo inference are presented in Section~\ref{sec:bayesian}, together with a simulation study that confirms the statistical identifiability of the model components.
Results for estimated arrival dates and other model components are presented for several species in Section~\ref{sec:results}. A discussion of results and an outlook towards future research related to our approach and  to citizen-science data more generally, concludes the paper in Section~\ref{sec:conclusion}, where we also highlight the potential utility of EVT in ecology.

\section{Datasets and preprocessing steps}\label{sec:data}

\subsection{Bird observation data}
\label{subsec:bird-data}
We extracted bird observation data from eBird and BBS databases for the study period  2001--2021. Both databases provide separate data files for each state in the selected study area in the northeastern US composed of the following states: Connecticut, Maine, Massachusetts, New Hampshire, New Jersey, New York, Pennsylvania, Rhode Island, Vermont. We merged the extracted data into a single dataset.  We extracted the full observation record of the eBird Basic Dataset, where each entry reports the observation of one or several individuals of a species, and retained the following attributes: name of species, count of observed individuals, longitude, latitude, date, duration of observation (i.e., temporal sampling effort), and a flag indicating if the observation is part of a so-called checklist. Recall that a checklist is an observation event where observers report all species they observe, provided they succeed in identifying the species. For BBS, we extracted information about observed species at the level of the survey route and year. 

The first arrivals of migratory birds in the study area are known to take place at the earliest towards the end of March. Here we identify migratory species as having no observations during the winter months of December and January in eBird. For the migratory species characterized by this property, we remove a few February occurrences that are understood as carryovers from the preceding autumn season.  Only migratory species that \REV{had at least 200 occurrences in eBird for 2021} and were also observed in BBS are kept. This selection procedure led us to consider around 50 species as migratory, and we apply our modelling approach to the ten species with the largest number of reported presences in eBird: Red-eyed Vireo, Eastern Wood-Pewee, Chimney Swift, Great Crested Flycatcher, Warbling Vireo, Veery, Chestnut-sided Warbler, Magnolia Warbler, Purple Martin, Blackburnian Warbler, in decreasing order of occurrence entries.

The following two data tables were generated from eBird:
\begin{itemize}
\item \emph{eBird checklists:} each row corresponds to one checklist in eBird, with attributes longitude, latitude, year and duration of observation;
\item \emph{eBird species occurrence:} for each  migratory species, the table contains the same number of rows as there are  checklists, and each row contains the attributes of the checklist, the name of the species, and the binary presence-absence flag to indicate if the species was observed. Species observations flagged as ``flyover" in eBird (i.e., where birds did not develop any nesting or breeding activity near the observation location) were declared as absences.  
\end{itemize}

We define four response variables for the components of our regression model based on BBS species counts and eBird data aggregated to a regular pixel grid of 20km width. This pixel size is similar to the spatial mapping unit used for most official eBird communications. Our response variables are:
\begin{itemize}
\item \emph{BBS species counts:} the number of observed birds, available for each combination of migratory species, route and year, where only routes surveyed during the specific year are considered; we use the Poisson distribution for this variable;  
\item \emph{eBird checklist count:} the number of available checklists, calculated for each configuration of year and pixel; we use the Poisson distribution for this variable;
\item \emph{eBird occurrence count:} the number of checklists (among all available checklists)  for which the species was reported present, calculated for each configuration of species, year and pixel; we use the binomial distribution for this variable;
\item \emph{eBird first arrival dates:} the minimum date of observation, expressed as the number of days since January 1st, calculated for each configuration of species, year and pixel where at least one observation of the species occurred; we use the Generalised Extreme-Value distribution (GEV) for the negated variable, which therefore represents a maximum. 
\end{itemize}
Some of the above datasets and their components are illustrated in Figure~\ref{fig:schematic}.

\subsection{Effect of observation effort on first arrival dates}

\begin{figure}[t]
    \centering
 \begin{subfigure}{0.23\textwidth}
\includegraphics[width=.99\textwidth]{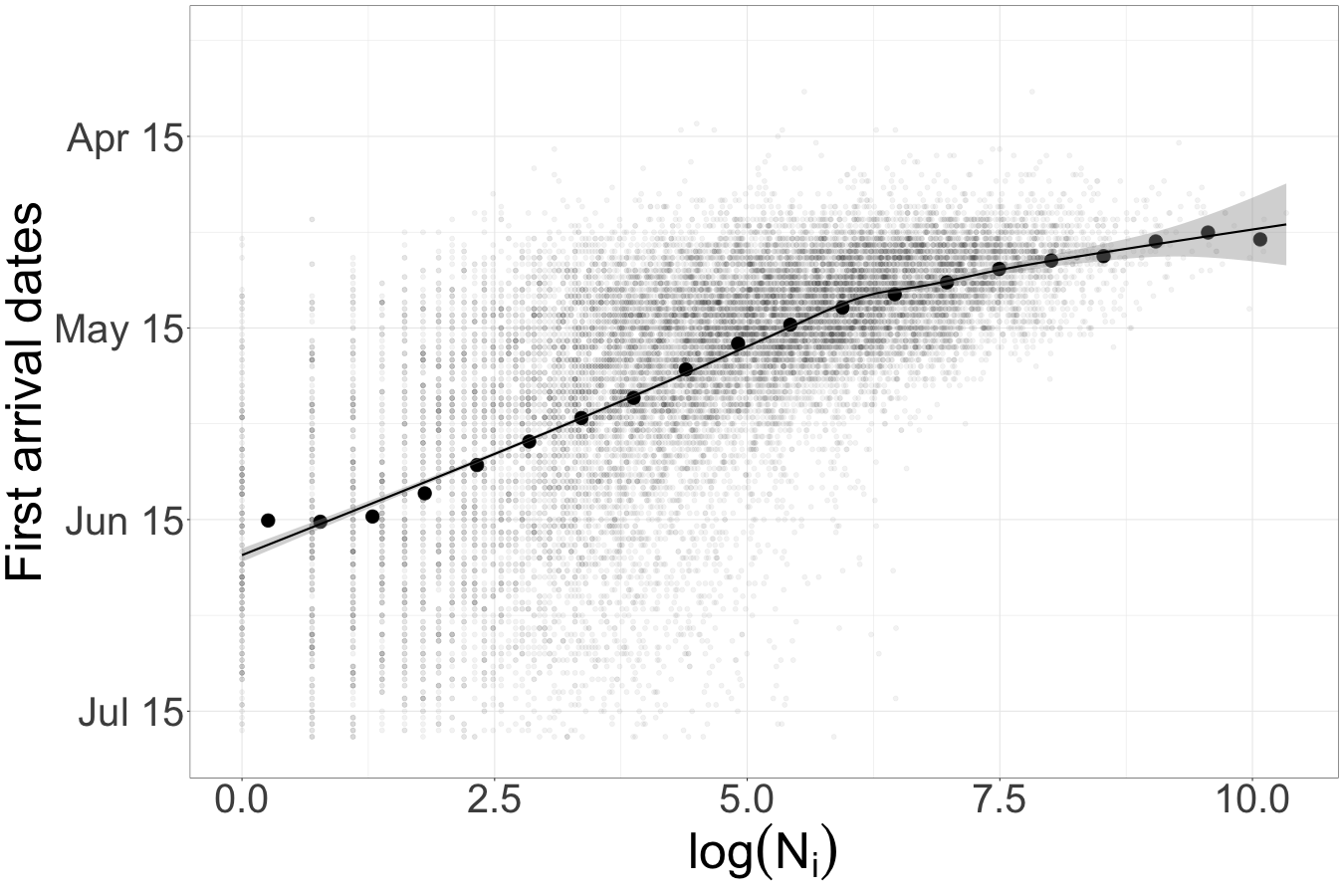}
\end{subfigure}  
 \begin{subfigure}{0.23\textwidth}
\includegraphics[width=.99\textwidth]{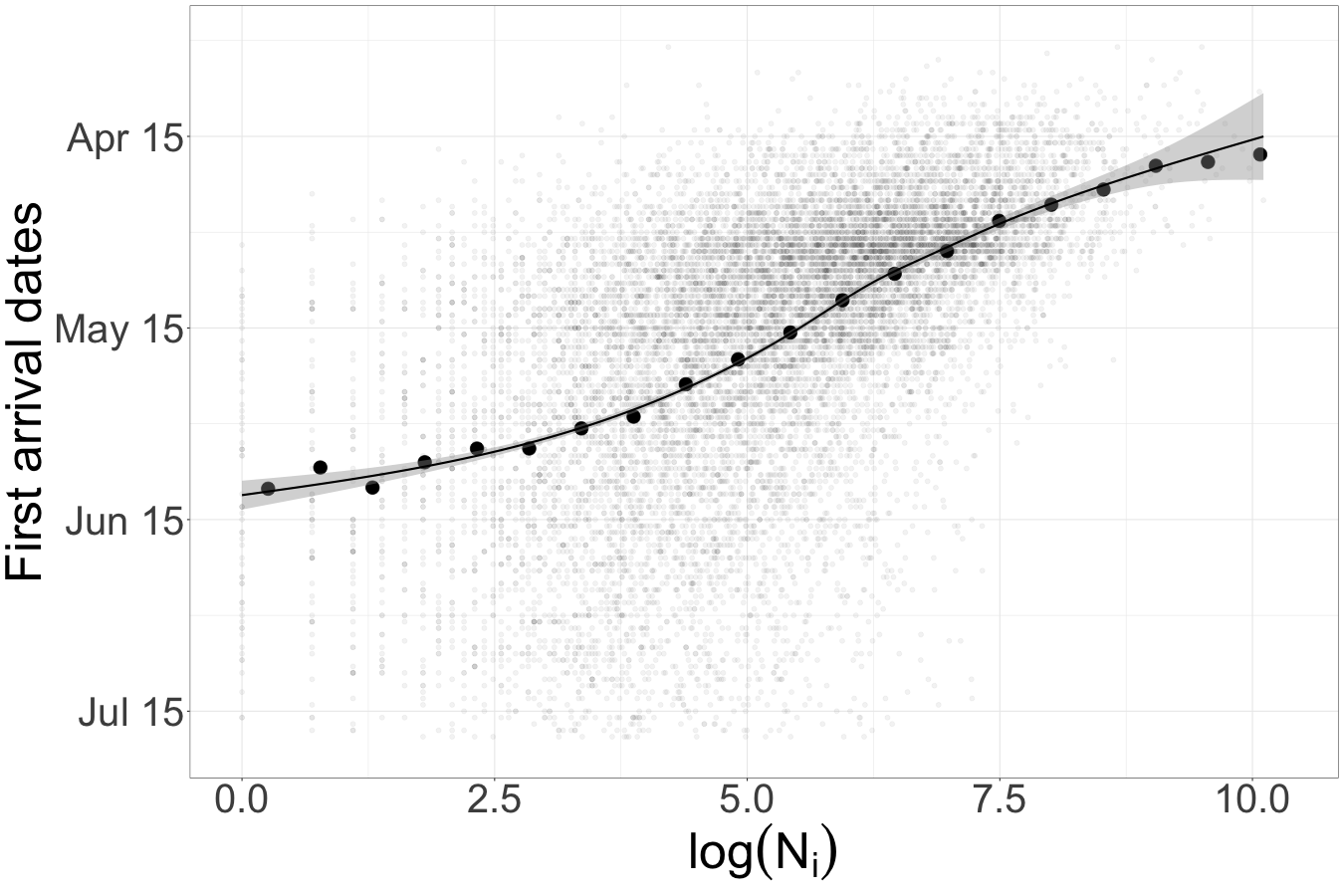}
\end{subfigure}  \\
\begin{subfigure}{0.23\textwidth}
\includegraphics[width=.99\textwidth]{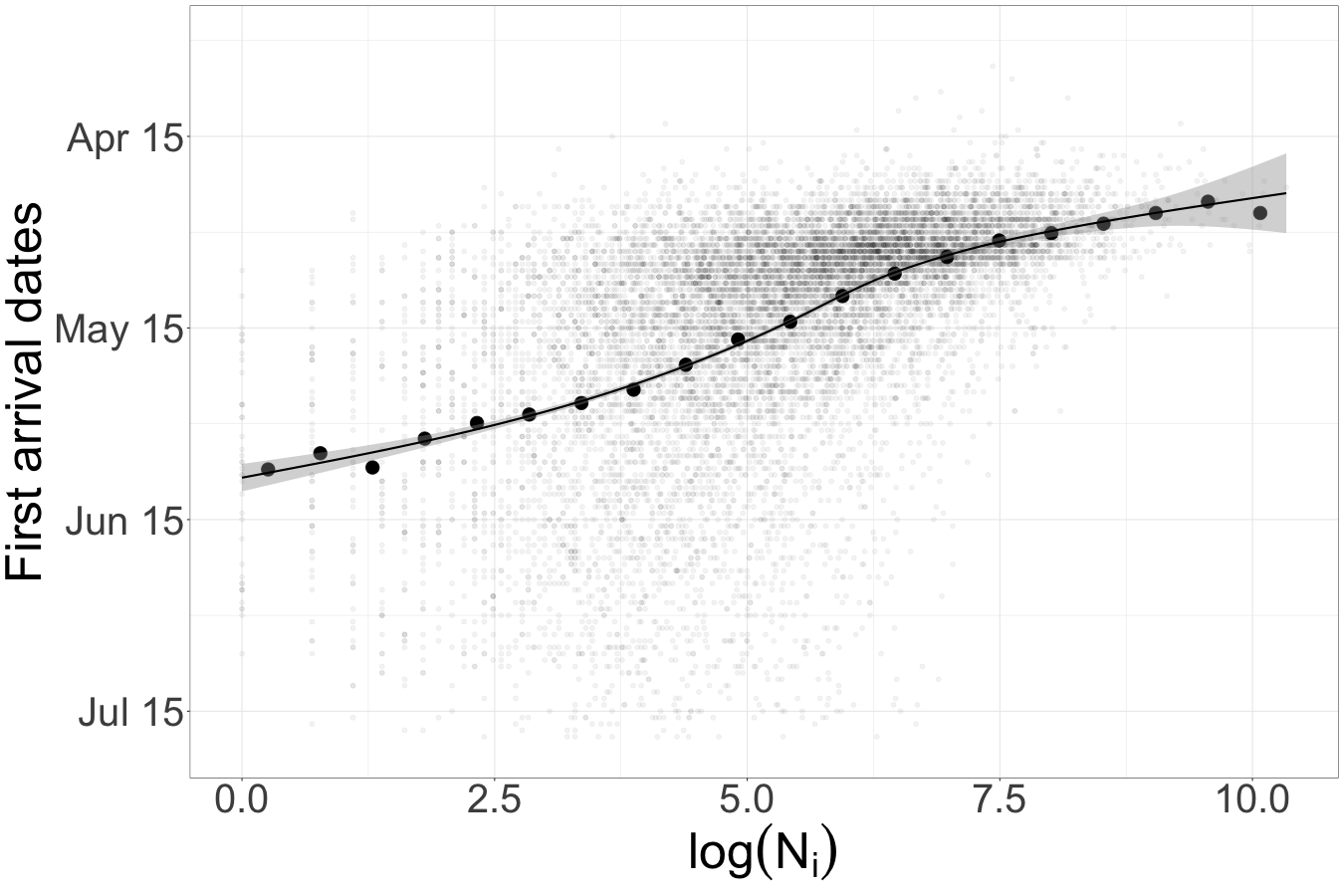}
\end{subfigure}
 \begin{subfigure}{0.23\textwidth}
\includegraphics[width=.99\textwidth]{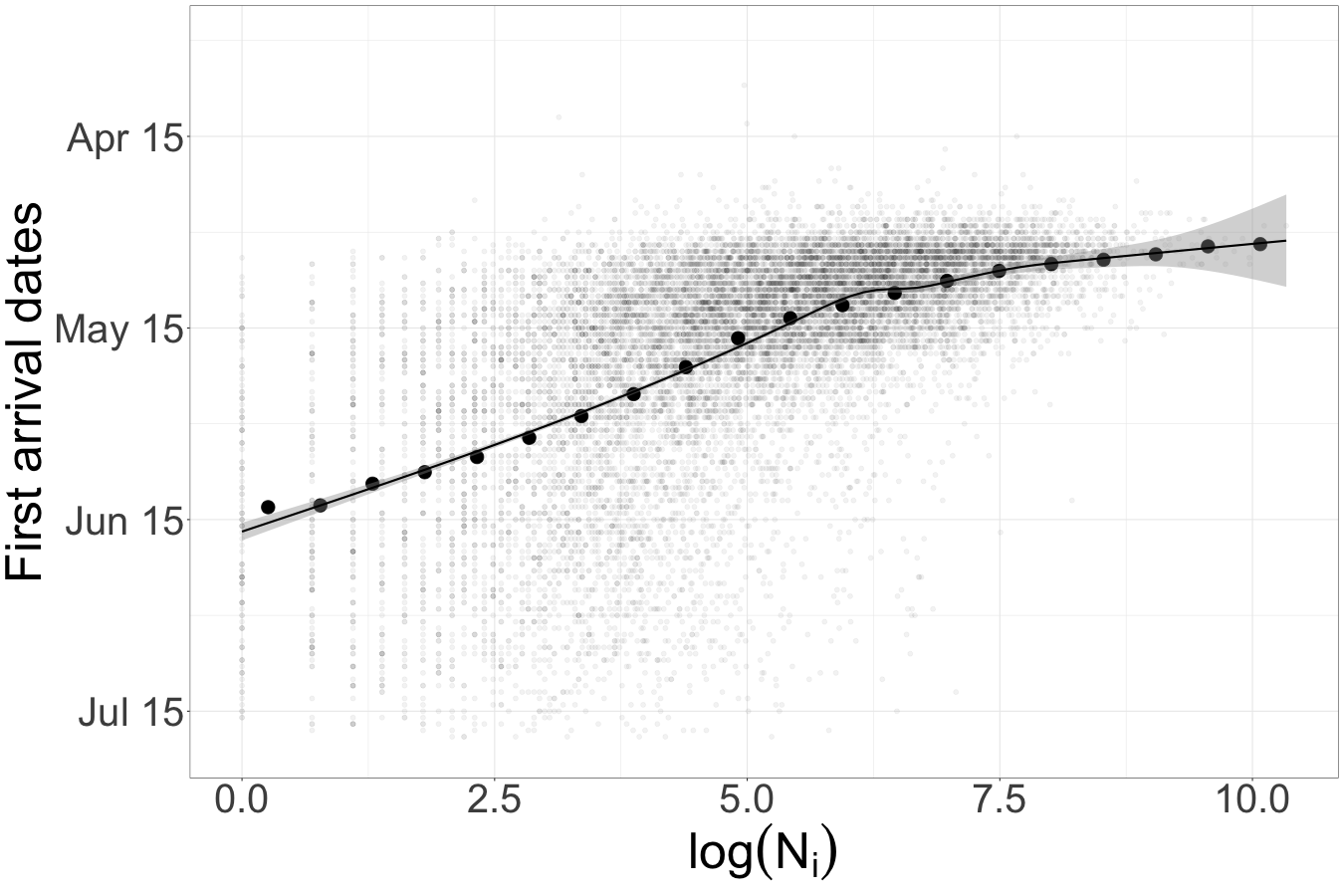}
\end{subfigure}  
  \caption{ Scatterplots of the logarithm of positive checklist counts (x-axes) and dates of first arrival (y-axes, with later dates corresponding to lower values) observed at each pixel-year  combination for the following species (top left to bottom right): \textit{Red-eyed Vireo}, \textit{Chimney Swift}, \textit{Warbling Vireo} and \textit{Chestnut-sided Warbler}. The larger dots show the 20 binned estimates, the black line a smooth fitted curve, and the shaded region the $95\%$ pointwise confidence intervals of the curve.}
  \label{fig:explore:preferential}
\end{figure}

Figure~\ref{fig:explore:preferential} shows observed first arrival dates plotted against positive checklist counts, for each pixel-year combination and several species. Observed dates tend to occur later during the year when fewer checklists are available, i.e., when sampling effort was lower in space. This provides us with strong motivation for modelling the influence of sampling effort on observed dates, and then using the estimated relationship to provide bias-corrected predictions for true dates by setting the sampling effort to a very high value during the prediction step.

\begin{figure}[t]
    \centering
 \begin{subfigure}{0.23\textwidth}
\includegraphics[width=.99\textwidth]{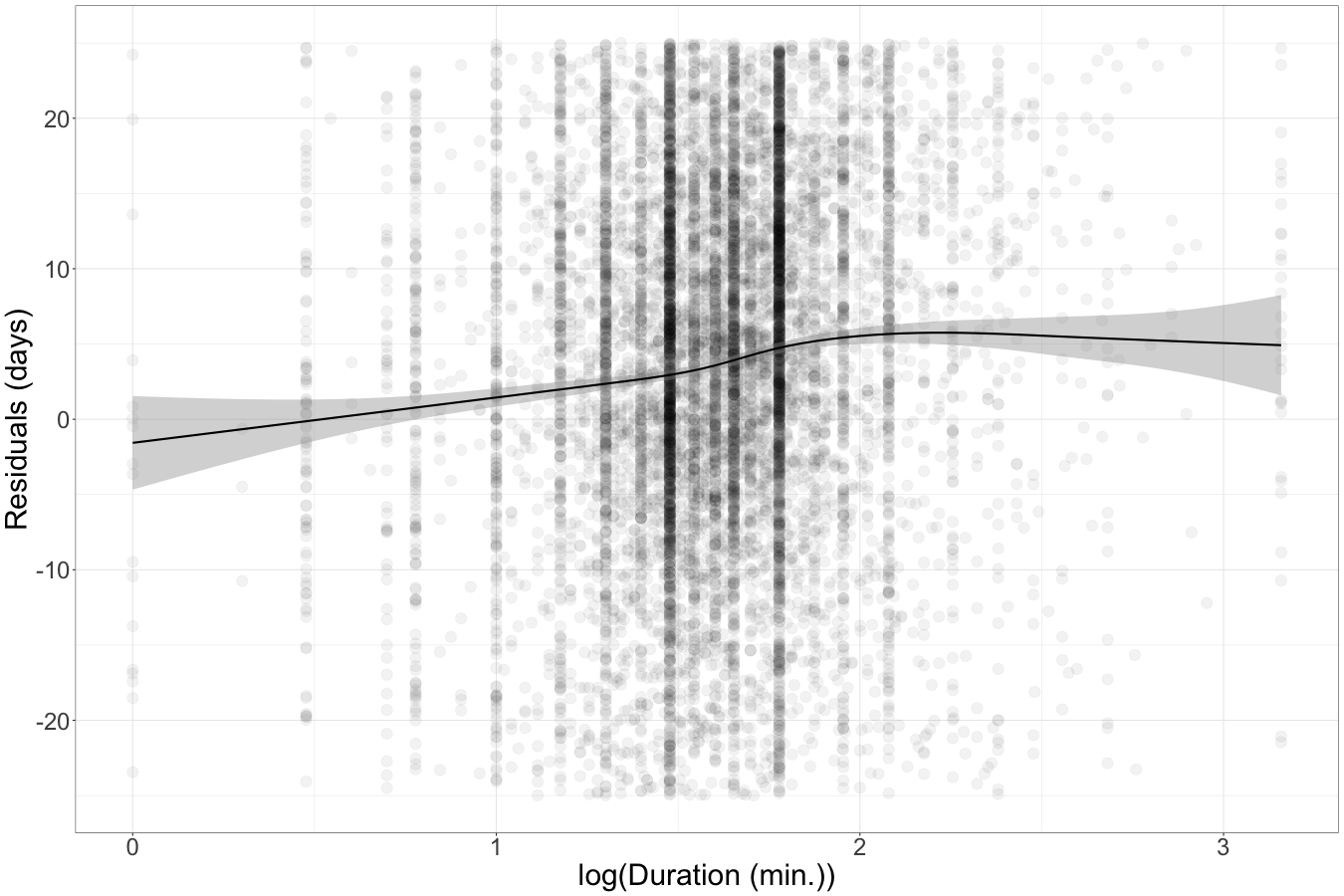}
\end{subfigure}  
 \begin{subfigure}{0.23\textwidth}
\includegraphics[width=.99\textwidth]{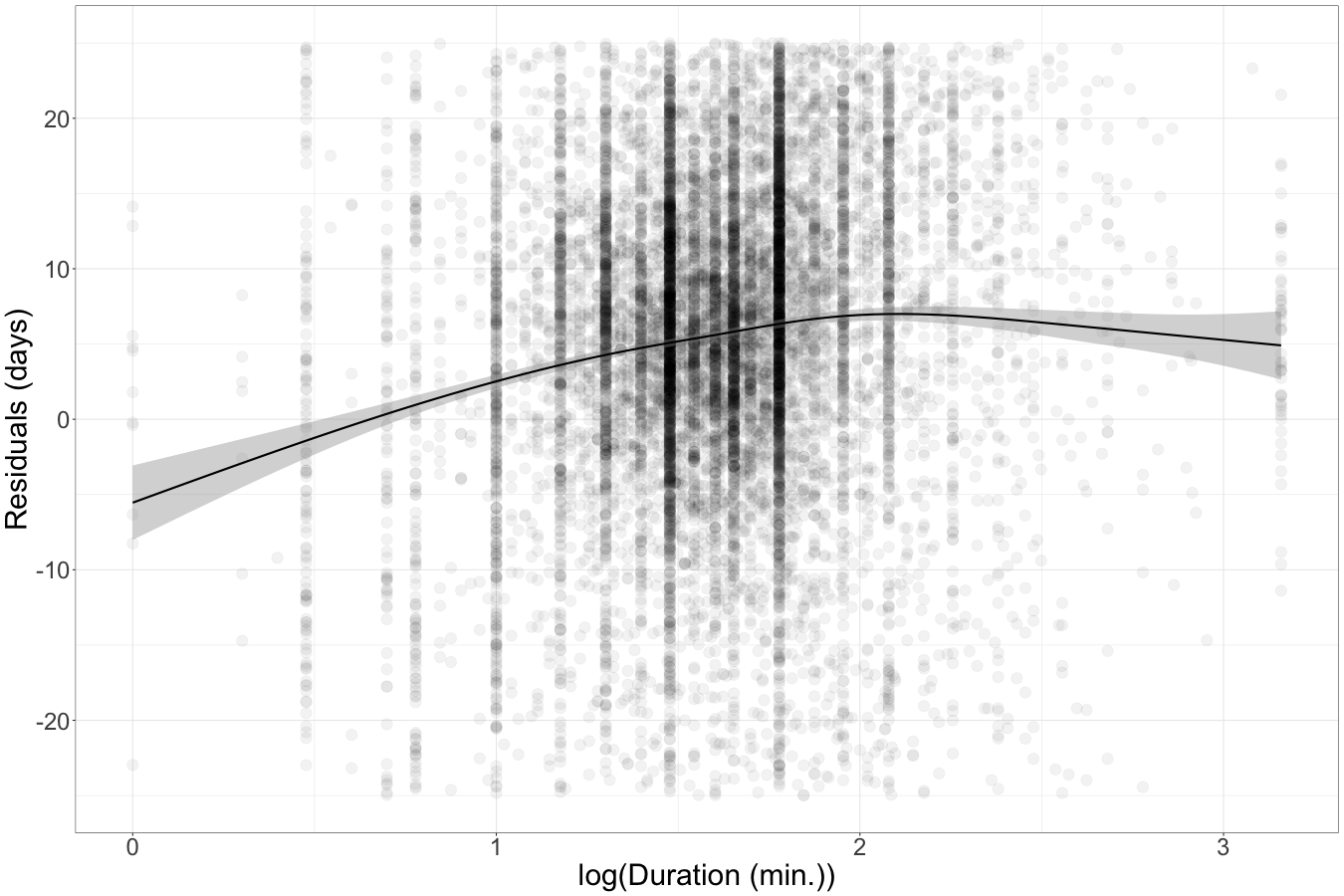}
\end{subfigure} 
  \caption{ Scatterplots of the inverse duration of observation during checklists (x-axis) and the residuals obtained from  fitting a smooth local regression curve to the dates of first arrival \REV{(with later dates corresponding to lower values)}  as a function of the logarithm (with base 10) of positive checklist counts (y-axis), for two species: \textit{Chimney Swift} (left); \textit{Chestnut-sided Warbler} (right). \REV{For improved readability of the figure,  the range of values shown for the residuals does not include all  points.}
  }
  \label{fig:explore:preferential2}
\end{figure}

Figure~\ref{fig:explore:preferential2} further explores whether the duration of observation during a checklist helps explain the residuals obtained after fitting a local regression curve in Figure~\ref{fig:explore:preferential}. We detect a slight effect of duration for the two example species in Figure~\ref{fig:explore:preferential2}\REV{, with relatively earlier arrivals arising for longer durations.} This suggests that both the number of checklists and the duration of observation per checklist are relevant components of the overall sampling effort modulating the distribution of the  observed first arrival date of birds at their breeding sites, with higher effort associated with earlier observation of the first arrival. 

To explain the remaining variability in the residuals  in Figures~\ref{fig:explore:preferential}~and~\ref{fig:explore:preferential2}, we conjecture that other factors related to spatial variation in bird phenology and in occupancy probabilities (i.e., the spatial ``range" of the species) are relevant. We construct our model to estimate spatial effects for these properties and then use them to correct predicted first arrival dates.

\subsection{Climate and land-cover data}

The North Atlantic Oscillation (NAO) Index describes changes in the strength of two recurring pressure patterns in the atmosphere over the North Atlantic: a low near Iceland, and a high near the Azores Islands. Positive NAO indicates these features are strong, creating a big pressure difference between them. Strongly positive values are linked to warm conditions across the Eastern US and Northern Europe, and cold conditions across Southern Europe. Negative NAO indicates these features are relatively weak, and the pressure difference between them is smaller. Strongly negative NAO is linked to cold conditions in the Eastern US and Northern Europe, and warm conditions in Southern Europe. As a covariate at annual temporal resolution, we use the average NAO for March, which was identified as a relevant climate predictor for bird migration timing by \citet{Wijeyakulasuriya.2023}. 

We use the 2021 National Land Cover Database of the US  to extract land-cover information. To facilitate model construction and identifiability, we will not use land-cover variables as covariates within our BHM. Instead, we will compare posterior maps of the niche, observation effort and components of the first arrival date with land cover maps to detect and interpret significant correlations between model components and land cover. 
Land cover proportions for the pixel grid are reported in the Appendix (Figure~\ref{fig:land-cover}) for four land-cover types,  defined by merging  original NLCD categories: Developed areas (including all areas with buildings and infrastructure such as roads); Forest; Vegetation (excluding forest but including planted and cultivated land); Water (including wetlands).

\begin{figure*}[t]
    \centering
\includegraphics[width=.99\textwidth]{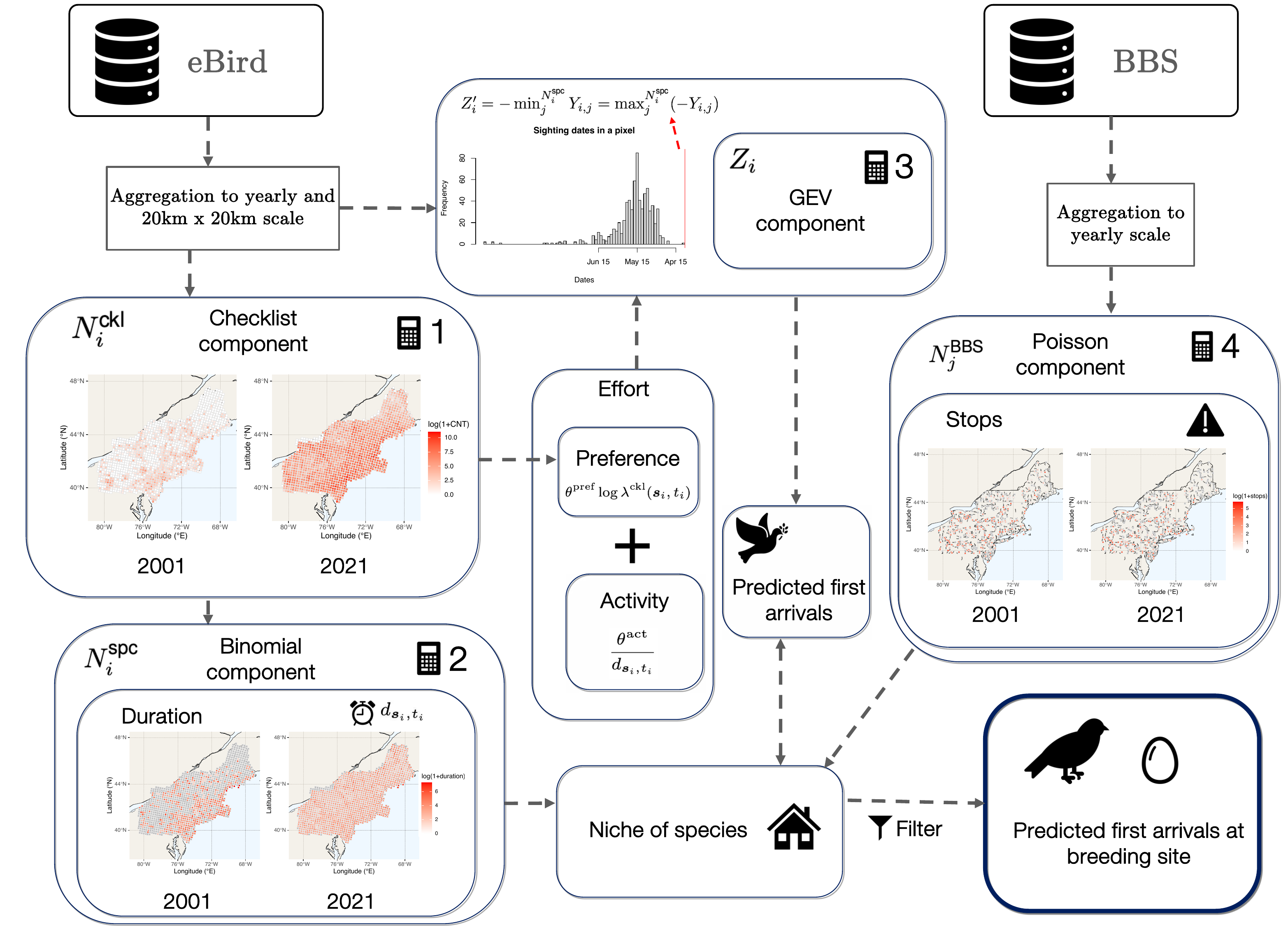}
  \caption{ Schematic summarising the data sources, the data preprocessing steps and the four components in the data layer of the BHM  (panels labelled with numbers 1 to 4). The dashed lines illustrate our modelling choices and how model components interact, but they need not show the chronology of how these components are estimated, since  the whole model is estimated jointly using a fully Bayesian approach. The plots in the  panels numbered `1', `2' and `4' show the number of available eBird checklists, the median duration spent per checklist in each pixel, and the number of stops along a route from the BBS data, respectively, in 2001 (left display in each panel) and 2021 (right display).}
  \label{fig:schematic}
\end{figure*}

Figure \ref{fig:schematic} shows a schematic summarising the different data sources and preprocessing steps of our approach, and the general workflow of the model. The plot in Panel~1 illustrates a strong increase in eBird checklists and in spatial coverage between the beginning and the end of the study period; the plot in Panel~2 highlights that the median duration per checklist varies substantially in both space and time. In fact, in 2001 we see longer durations in specific spatial regions, while in 2021 the durations are more homogeneous across larger areas. There is also a general overall increase in activity. The display in Panel~3 shows the occurrence dates of  observations of a species pooled together for a pixel-year, and we extract the minimum day of the year, which will serve as an observation for the GEV response variable in our BHM. Finally, the figure in Panel~4 shows the number of BBS stops in 2001 and 2021.

\section{Bayesian hierarchical model}\label{sec:model}

\subsection{General setting and notations}

Figure~\ref{fig:schematic} also highlights the four components in the data layer of the BHM we propose, and we here detail its notations and structure.

We write $\mathcal{S}\times \mathcal{T}$ for the space-time study domain, and denote by $A_{i}$, $i=1,\ldots,D\times T$, the set of non-intersecting space-time cells based on a division of the study area $\mathcal{S}$ into $D=1268$ pixels, replicated over $T=21$ years. We identify each cell $A_i$ with a representative location $(\bm s_i,t_i)$, such as its barycenter, where $\bm s_i \in \mathcal{S}$ and $t_i=1,\dots,T$.

Let $N^\text{ckl}_i$ be the number of eBird checklists in $A_i$. If $N^\text{ckl}_i>0$, then $N^\text{spc}_i\in\{0,1,\ldots,N^\text{ckl}_i\}$ denotes the sum of the binary indicators detecting a given species from each of the $N^\text{ckl}_i$ checklists. Binary indicators take a value of $1$ if the species was detected and $0$ otherwise. Given $N^\text{spc}_i>0$, we write $\bm Y_{i}=(Y_{i,1},\ldots,Y_{i,\REV{N^\text{spc}_{i}}}) \in [1,365]^{N^\text{spc}_{i}}$ to denote the corresponding arrival dates in $\bm s_i$ within the year $t_i$ from the $N^\text{spc}_i$ observed species occurrences. Based on these arrival dates, we calculate $Z^\prime_i=-\min_{j}^{\REV{N^\text{spc}_i}} Y_{i,j} = \REV{\max_{j}^{N^\text{spc}_i} (-Y_{i,j})}$, the first arrival date in  year $t_i$ within $\bm s_i$. We set $Z_i=\REV{-\log(-Z_i^\prime/366)}$ to reflect that $Z_i^\prime$ has a natural lower bound at $-366$, and model \REV{$Z_i >0$} instead. 




\subsection{Extreme-value distribution for first arrival dates}

Provided that the partitioning of $\mathcal{S}$ provides enough `\REV{effectively} independent' arrival times within each pixel-year $A_i$, Extreme-Value Theory \citep[EVT,][]{Coles.2001} motivates using a Generalised Extreme-Value distribution (GEV) to model  the transformed first arrival date in $A_i$. Invoking EVT, we assume convergence in distribution of each $Z_i$ to a limiting GEV as $N_i^{\text{spc}}\rightarrow\infty$. 
More precisely, let $M_n$ denote the maxima of independent and identically distributed random variables $X_1,\dots, X_n$. If $M_n$ converges in distribution to a non-degenerate distribution after linear renormalisation, then this distribution must be a GEV\REV{; this result was further extended and shown to hold for maxima over dependent data, although under technical assumptions that are difficult to validate in applications}. In practice, for fixed $n$ large enough (where $n=\REV{N_i^{\text{spc}}}$ in our setting), one uses the approximation
	\begin{equation}
	\label{eqn:gevcdf}
	\text{Pr}(M_n \leq z) \approx \begin{cases}
		\exp\left\{-\exp\left(-\frac{z-\mu}{\sigma}\right)\right\}, &\xi = 0,\\[5pt]
		\exp\left\{-\left[1 + \xi\left(\frac{z-\mu}{\sigma}\right)\right]^{-1/\xi}\right\}, &\xi \neq 0,
	\end{cases}
	\end{equation}
 where the right-hand side of \eqref{eqn:gevcdf} is the distribution function of the GEV, defined on $\{z:1+\xi(z-\mu)/\sigma>0\}$ with location $\mu\in\mathbb{R}$, scale $\sigma>0$ and shape $\xi \in \mathbb{R}$. The shape parameter determines the tail behaviour of the GEV: for $\xi>0$, the GEV has support with a finite lower endpoint and  an upper heavy tail of power-law form; for $\xi<0$, the GEV has a finite upper endpoint but is unbounded below; for $\xi=0$, the support of the GEV has no finite bounds, and the upper tail exhibits an exponential decay rate. In our application, some sample sizes \REV{$N_i^{\text{spc}}$} can be very small and dates $Y_{i,j}$, $j=1,\ldots,\REV{N_i^{\text{spc}}}$ can be dependent, but the three-parameter  GEV remains a flexible model to accommodate relevant distributional properties of the maxima $Z_i$, even if the theoretical asymptotics are not always reached. 
\REV{The transformation from $Z_i'$ to $Z_i$ is very close to linear in a large neighbourhood of $Z_i'=100$ (corresponding to approximately mid-April), so it does not substantially modify the shape $\xi$ but the parameters $\mu$ and $\sigma$.}  
We incorporate linear predictors into the GEV regression equation used to model $Z_i$ in our BHM: one for the location $\mu$ and another for the scale $\sigma$. 

\subsection{Data layer of the Bayesian hierarchical model}\label{subsec:model:hierarchical}



We construct a system of four regression equations with latent Gaussian processes used in the linear predictors. We denote the response variables, as described in Section~\ref{subsec:bird-data}, as follows: 
$N_i^{\text{ckl}}$  -- eBird checklist count per pixel-year \REV{$i$}; $N_j^{\text{BBS}}$ -- BBS counts of the species of interest per year-route $j$; $N_i^{\text{spc}}$  -- eBird occurrence count for the species of interest per pixel-year \REV{$i$}; $Z_i$  -- transformed negated first arrival date for the species of interest in eBird per pixel-year $i$.

For the response variable $N_i^{\text{ckl}}$ reporting the checklist counts, we weight the Poisson intensities according to the surface areas \REV{$\mathcal{A}_i$} of the pixels (i.e., we use an offset of \REV{$\log \mathcal{A}_i$} in the linear predictor), since some pixels at the boundary of the study region have smaller area. For the response variable $N_j^{\text{BBS}}$ reporting the BBS occurrence counts, observations are made along a subset of 50 equidistant stops  along each of the routes, and we weight the Poisson intensities accordingly with the number of stops (between $1$ and $50$) that were visited. To alleviate notations in the predictor formulas below, we do not explicitly write down these weights related to pixel surface areas and road lengths. 
Moreover, routes can span across several pixels, so we construct the Poisson intensity parameters as a weighted mean of the parameters for the involved pixels with weights $w_k$ defined as the percentage length inside each of the pixels $A_k$. 
    
The hierarchical system of the four regression equations is
\begin{align}
N^{\text{BBS}}_j \mid \lambda^{\text{BBS}},\bm \theta_\text{bbs}  &\sim \mathrm{Pois}\left\{ 
\sum_{k\in \text{route}_j}\omega_k\lambda^{\text{BBS}}(\bm s_k; \bm \theta_\text{bbs} ) \right\}, \nonumber \\	
	N^{\text{ckl}}_i \mid \lambda^{\text{ckl}}, \bm \theta_\text{ckl} &\sim \mathrm{Pois}\left\{\lambda^{\text{ckl}}(\bm s_i, t_i ; \bm \theta_\text{ckl}) \right\}, \nonumber\\ 
   N^\text{spc}_i \mid \REV{N^{\text{ckl}}_i}, p^\text{spc}, \bm \theta_\text{spc} &\sim \mathrm{Bin}\{N^{\text{ckl}}_i, p^\text{spc}(\bm s_i, t_i; \bm \theta_\text{spc}) \}, \nonumber\\
	Z_i\mid \mu, \REV{\bm \theta_\mu}, \sigma, \bm \theta_\sigma  &\sim \mathrm{GEV}\{ \mu(\bm s_i, t_i; \bm \theta_\mu), \sigma(\bm s_i; \bm \theta_\sigma ), \xi \}, \nonumber
\end{align}
where
$$
\bm \theta_\text{bbs}, \bm \theta_\text{ckl}, \bm \theta_\text{spc}, \bm \theta_\mu, \bm \theta_\sigma \sim \text{Hyperpriors}
$$
are hyperparameters for the random predictors $\lambda^{\text{BBS}}$, $\lambda^{\text{ckl}}$, $p^\text{spc}$, $\mu$ and $\sigma$ that govern the different model components; we discuss their specifics next. Checklist occurrences form a point pattern, and due  to the structure of the $N_i^{\text{ckl}}$-related model component,  we model them as a spatiotemporal log-Gaussian Cox point process \citep{Moller.al.1998}, discretised according to the pixels $A_i$. 

\subsection{Structure of latent Gaussian processes}\label{subsec:model:spatial}

We write $X(\bigcdot)$ to denote any latent Gaussian effect, here either indexed by pixels $\bm s\in\mathcal{S}$ or years $t\in\mathcal{T}$. We use Gaussian process priors for all  latent effects. We further use the exponential covariance function to parametrise the spatial effects, given as follows for two locations $\bm s_1$ and $\bm s_2$:
$$
\mathrm{Cov}\{X( \bm s_1), X( \bm s_2)\} = \sigma^2 \exp(- ||  \bm s_1 -  \bm s_2||/\kappa), \quad \sigma, \kappa >0,
$$
with Euclidean distance $||\cdot||$, standard deviation $\sigma>0$ and range parameter $\kappa>0$. 

The four spatial Gaussian process priors included in our model are 
\begin{align}
\REV{X^{\text{pref}}(\bigcdot)} \sim  \mathcal{GP}(\bm \omega_1), \quad X^{\text{niche}}( \bigcdot)  \sim  \mathcal{GP}(\bm \omega_2), \notag\\
X^{\text{GEV-}\mu}( \bigcdot) \sim  \mathcal{GP}(\bm \omega_3), \quad 
X^{\text{GEV-}\sigma}( \bigcdot)  \sim  \mathcal{GP}(\bm \omega_4), 
\label{eq:latent-fields}
\end{align}
where $\bm \omega_1$, $\bm \omega_2$, $\bm \omega_3$ and $\bm \omega_4$ contain individual range and standard deviation parameters that control the spatial dependence of each process. These hyperparameters are assigned identical joint Penalised Complexity priors \citep{Fuglstad.al.2018}. We also impose sum-to-zero constraints on all spatial effects to aid with identifiability. Spatial patterns of the niche of the species are captured by $X^{\mathrm{niche}}$, whereas spatial patterns of the sampling effort are modelled through \REV{$X^{\mathrm{pref}}$}. \REV{Moreover, we include a temporal latent effect $X^\text{year}$ to capture the variability in checklist counts across years, and set the prior  $X^\text{year}(\cdot) \sim \mathcal{GP}(\bm \omega_5)$ \REV{where $\bm \omega_5$ contains the range and standard deviation parameters of a temporal exponential correlation function}.}

The structure of link functions and linear predictors, explained in detail subsequently, is:
\begin{align*}
\log \lambda^\text{BBS}(\bm s_i) &= \beta_0^\text{BBS} + X^\text{niche}(\bm s_i),\\
 \log \lambda^{\text{ckl}}(\bm s_i, t_i) &= \beta_0^{\text{ckl}}  +  X^\text{year}(t_i) + X^{\text{pref}}(\bm s_i), \\
 \mathrm{cloglog}\{\REV{p^\text{spc}}(\bm s_i,t_i) \} &= \beta_0^{\text{spc}} + X^{\text{niche}}(\bm s_i)   +
    \dfrac{\REV{\beta^{\text{act}}} }{ d_{\bm s_i, t_i} },\\
     \mu(\bm s_i, t_i) &=     g \{\beta_0^{\text{GEV-}\mu}+X^{\text{GEV-}\mu}(\bm s_i) + \beta_1^{\text{GEV-}\mu} \mathrm{NAO_{t_i}} \\ & \quad\quad  + \theta^\text{niche-GEV} X^{\text{niche}}(\bm s_i) , 
 x_\text{effort}(\bm s_i, t_i)\}, \\
\log \sigma(\bm s_i) &= \beta_0^{\text{GEV-}\sigma} + X^{\text{GEV-}\sigma}(\bm s_i). 
\end{align*}
The climate covariable, i.e. the NAO value for year $t_i$, is denoted as $\mathrm{NAO_{t_i}}$, and we include it inside the link function $g$ (detailed below) as a fixed effect modulating the upper bound of the GEV location parameter.
\REV{The scaling parameter  $\theta^\text{niche-GEV}$ controls} the scale of sharing of the niche spatial effect. \REV{All scaling hyperparameters and intercepts in the model, along with $\xi$, are given flat normal prior distributions with mean $0$ and variance $100$.}  The complementary log-log link function $\text{cloglog}(p)=\log(-\log(1-p))$ for binomial probabilities allows interpreting them as the probabilities  of observing at least one count in a Poisson model (which is equal to $1-\exp(-\lambda))$ with Poisson intensity \REV{$\lambda=\exp(x)$}. This ensures consistency with the Poisson model for the BBS counts. 

We differentiate between two dimensions of sampling effort that we combine into $x_\text{effort}$. The first one is modelled through the space-time intensity  $\lambda^\text{ckl}$ of checklists, which we call the \textit{preference}. We hypothesise that it influences  the first arrival dates $Z_i$, where a higher \textit{preference} likely leads to more and thus earlier sightings in the year. 
The second dimension of sampling effort that we consider here is  the median duration of observation per checklist in a pixel-year $d_{\bm s_i, t_i}$, 
which we call the \textit{activity}. This effect could influence both the binomial probability of observing the species of interest in a checklist, and again the first arrival dates. A higher \textit{activity} could increase the probability of observing the species in a checklist, which could also induce an earlier observed arrival date.  We  combine the two effects \textit{preference} and \textit{activity} into 
\begin{equation}\label{eq:x-effort}
x_\text{effort}(\bm s_i, t_i) = \REV{\theta^\text{eff}+} \theta^\text{pref} \log \lambda^{\text{ckl}}(\bm s_i, t_i)+ \dfrac{\theta^\text{act} }{ d_{\bm s_i, t_i} }
\end{equation}
with \REV{real-valued} scaling hyperparameters $\theta^\text{pref}$ \REV{(typically expected to be positive)} and $\theta^\text{act}$ \REV{(typically expected to be negative)}\REV{, and an intercept $\theta^\text{eff}$ representing a baseline sampling effort that is modulated according to the checklist abundance and observer activity}.  With this structure, setting an infinite duration $d_{\bm s_i, t_i}$
causes the activity term to vanish. 

The next subsection explains the choice we make for the nonlinear link function $g$.

\subsection{Saturated sharing towards the GEV predictor}


Sharing latent processes across carefully chosen regression equations, as implemented in Section~\ref{subsec:model:spatial}, enables us to leverage data from several response variables 
to identify the four latent processes. Specifically, we want to identify the  effect of spatiotemporally heterogeneous sampling effort and correct it in predictions, such as those of the true first arrival dates. 

\begin{figure}[t]
    \centering
 \begin{subfigure}{0.35\textwidth}
\includegraphics[width=.99\textwidth]{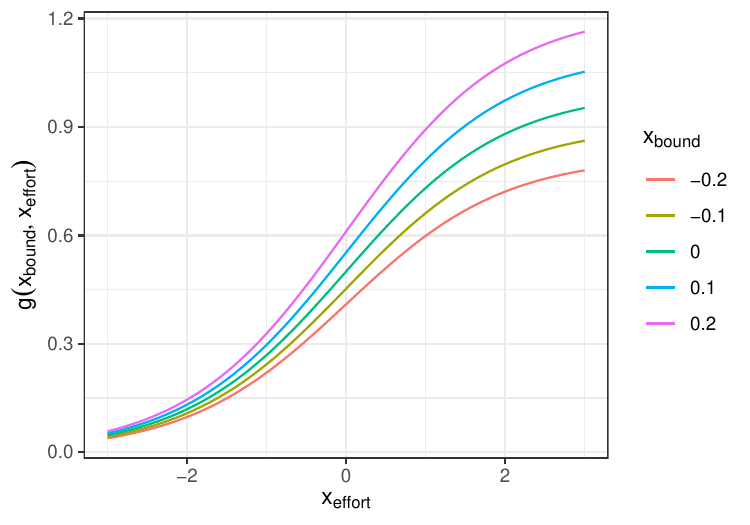}
\end{subfigure} \\
  \caption{ Plots of the nonlinear sharing function $g$ in \eqref{eq:g_function} for varying  $x_\text{effort}$ (x-axis) and $x_\text{bound}$ (colours). 
  }
  \label{fig:g_func}
\end{figure}

The most often used approach of sharing latent effects between regression equations is via a linear structure where the linear predictor of one regression equation has an additive component $X(\bigcdot)$, while another has an additive component $\beta \times X(\bigcdot)$, with a scaling factor $\beta\in\mathbb{R}$. This yields models akin to the preferential-sampling framework of \citet{Diggle2010} and \citet{Pati.2011}, and to \citet{Koh.2022} and \citet{yadav.2022} in the context of joint modelling  of occurrences and sizes of wildfires or landslides. However, such linear sharing is not flexible enough for sharing the sampling effort towards the linear predictor of the $\mu$  parameter of the GEV for first arrival dates. Letting the effort tend to infinity with linear sharing would lead to first arrival dates tending to January 1st, the smallest possible value. 
As illustrated by the exploratory data analysis in Figure~\ref{fig:explore:preferential}, the observation effort effect on first arrival dates saturates at very high levels. 
Therefore, we share $x_\text{effort}$ with the $\mu$ component of the GEV  through the following upper-bounded function 
\begin{equation}\label{eq:g_function}
g(x_\text{bound},\,x_\text{effort}) =  \frac{\REV{\exp(x_\text{bound})}}{1+\exp(-x_\text{effort})},
\end{equation}
\REV{where the first argument determines the upper bound in the GEV location parameter, which is modulated by the sampling effort in the second argument and can be attained with unlimited (i.e., infinite) effort. The $g$-function is always positive, which is appropriate here when considering as response the transformed dates $Z_i>0$. }  
Figure~\ref{fig:g_func} illustrates that the class of functions \eqref{eq:g_function} is bounded and monotonically increasing, therefore allowing the effect of the sampling effort on the first arrival dates to saturate with high effort. 

The structure we put into \eqref{eq:g_function} is also inspired by species accumulation curves in ecology \citep{Colwell.1994}, but here we replace the species abundance with the first observed arrival dates. To our knowledge, such an approach has not yet been considered in the literature. \REV{It establishes a representation that is both parsimonious and ``physically meaningful", a feature that is desirable in statistical modelling in various contexts such as landslide prediction in \citet{Opitz2022} or temperature extremes in \citet{Noyelle2024}.}


\section{Simulation-based Bayesian inference}\label{sec:bayesian}

Fast and relatively accurate off-the-shelf implementations of Bayesian hierarchical inference exist, such as the INLA framework \citep{Rue.al.2017}. However, due to certain specifics of our model, such as the nonlinear sharing of latent spatial effects and the inclusion of linear predictors in several parameters of the GEV, it is not feasible to use them here. Instead, we develop a Markov-Chain Monte-Carlo sampler for simulation-based inference. 

\subsection{Vecchia approximation for Gaussian components}\label{subsec:vecchia}


Due to the large number of spatial cells in our case study ($D=1268$), keeping spatial  latent Gaussian effects and their conditional distributions numerically tractable is challenging. Various solutions have been proposed to tackle this fundamental problem in  spatial statistics with large-dimensional data, including  Gauss--Markov random fields used to approximate certain types of covariance functions based on Stochastic Partial Differential Equations \citep[SPDE approach,]{Lindgren.al.2011}, or Nearest-Neighbor Gaussian processes \citep[NNGPs,][]{Datta2016}. 
Here we use another flexible and popular option to construct approximate Gauss--Markov representations of any Gaussian covariance function through the \citet{Vecchia.1988} approximation, which  \REV{leads to a sparse Cholesky factor} of the precision matrix.  \REV{We write $\{x_1, \dots, x_D\}$ for the set of observations of a Gaussian field $X$ evaluated at locations $s_1,\dots, s_D\in\mathcal{S}$, and we consider a  permutation $m:\{1,\dots,D\} \rightarrow \{1,\dots,D\}$ defining a reordering of the observations. Based on this reordering, we can write $H(i;m) = \{ j \in \{1,\dots,D\} : m(j) < m(i) \}$ for the ``history'' of the $i^\text{th}$ index based on the permutation $m$, with $\bm x_{H(i;m)}$ denoting the corresponding subvector of observations. For a chosen permutation $m$, the exact joint density of the observations can be written as the product of conditional densities, i.e.,  
$$
f(x_1, \dots, x_D) = f(x_{m(1)}) \prod^D_{i=2}f(x_{m(i)}\mid \bm x_{H(i;m)}).
$$
The Vecchia approximation still represents a valid Gaussian process, but assumes that 
\begin{align}\label{eq:vecchia}
    f(x_1, \dots, x_D) &\approx \hat{f}(x_1, \dots, x_D) \nonumber \\
    &= f(x_{m(1)}) \prod^D_{i=2}  f(x_{m(i)}\mid \bm x_{S(i;m)}), 
\end{align}
where $S(i;m)\subseteq H(i;m)$ and $|S(i;m)| = k$. The approximation} deliberately reduces the conditioning history in the conditional densities, which decreases the computational complexity of evaluating the density from $\mathcal{O}(D^3)$ to $\mathcal{O}(Dk^3)$. This reduction is considerable if $k\ll D$, as in our setting  with the number of spatial pixels $D>1000$. An active research area \citep{Katzfuss.Guinness.2021} is to understand how one should \REV{choose the permutation $m$ and} order the conditioning variables in \eqref{eq:vecchia} to obtain approximations that satisfy certain criteria of optimality. \citet{Vecchia.1988} suggested to order the conditioning variables lexicographically based on their spatial coordinates, but this approach has been shown to be inefficient \citep{Guinness.2018}. Instead, we follow the ordering proposed by the latter study, which ensures that points are ordered in a quasi-random fashion. We select $k=5$ conditioning locations in our case study to make inference feasible. The simulation study detailed in Section \ref{sec:sim} confirms that this is reasonable.

\subsection{Proposal scheme for Markov-Chain Monte-Carlo}



We draw posterior inference on the parameters using a general Markov-Chain Monte-Carlo (MCMC) sampling scheme with customised Metropolis-Hastings (MH) updates. The hyperparameters are updated using Gibbs sampling. The  most computationally intensive part of our scheme involves sampling from the latent Gaussian components, which are updated jointly through a Metropolis Adjusted Langevin Algorithm \citep[MALA;][]{Roberts.Rosenthal.2001} designed to speed up mixing of the chains. We write $\bm x$ for the \REV{current state of the} latent Gaussian component of interest, $\bm x^\star$ for a proposal for that component, and $\bm y$ for the data vector. A discrete approximation of the Langevin diffusion implies that MALA is a MH algorithm with proposal distribution 
\begin{equation}\label{eq:langevin}
    ( \bm x^\star | \bm x)  \sim N_p \left(\bm x + \dfrac{\delta^2}{2} \nabla_{\bm \theta} \log \pi(\bm x | \bm y), \delta^2 \bf{W}_p \right),
\end{equation}
where tuning parameters are the pre-whitening matrix $\bf{W}_p$, used to account for the correlation between parameters within the component, and the stepsize $\delta$. For example, if we set $\bm x = \{X^\text{niche}(\bm s_1),\dots, X^\text{niche} (\bm s_D)\}$, the gradient in \eqref{eq:langevin}, when updating this component, is 
\begin{align*}
    \nabla_{\bm x} \log \pi(\bm x | \bm y) \approx &\nabla_{\bm x} \log \pi_\text{BBS}(\bm N^\text{BBS} \mid \bm x) 
    + \nabla_{\bm x} \log \pi_\text{spc}(\bm N^\text{spc} \mid \bm x) \\ 
    &+ \nabla_{\bm x} \log \pi_{\text{GEV}}(\bm Z \mid \bm x)
    + \nabla_{\bm x} \log \pi_\text{Vecchia}(\bm x),
\end{align*}
where the first three terms represent the gradients of log-likelihoods for the model components that $X^\text{niche}$ is included in, and  
$$
\nabla_{\bm x} \log \pi_\text{Vecchia}(\bm x) = -\Tilde{\Sigma}_{\bm\omega_2}^{-1} \bm x,
$$
where the $D\times D$ precision matrix $\Tilde{\Sigma}_{\bm\omega_2}^{-1}$ is the Vecchia-approximated precision of the Gaussian field prior assigned to $X^\text{niche}$, as detailed in \S\ref{subsec:vecchia} and implied by \eqref{eq:vecchia}. The precision matrix is parametrised by the vector $\bm\omega_2 =(\sigma, \kappa)$, see \S\ref{subsec:model:spatial}, to which we assign a hyperprior.

For our data application, we parallelise and run separate chains for each species. We generate \REV{$80'000$} posterior samples and discard the first $60'000$ as the burn-in period. We then perform a standard thinning operation of the Markov chains and keep one of four consecutive samples, thus obtaining $5'000$ samples overall to perform posterior inference. We divide the number of posterior samples and the burn-in period by two for our simulation study for computational reasons. We monitor the convergence and mixing of the chains through trace plots and by assessing the effective sample size. Code implemented  
for the MCMC procedure is available at \url{https://github.com/kohrrelation/mcmc_birds}. To obtain a shorter burn-in period and to avoid having  initial values too far from the region where posterior mass concentrates, we ran test chains that model each model component separately first (where we can estimate all parameters except the sharing parameters).  \REV{When possible},  we then used the last values of those chains as starting values when fitting the full model. Calculations were performed on UBELIX\footnote{\url{https://www.id.unibe.ch/hpc}}, the HPC cluster at the University of Bern, and the computational time of our algorithm there is roughly 6 seconds per iteration. 


\subsection{Simulation study }\label{sec:sim}

We check by simulation that the model parameters 
are identifiable and can be estimated appropriately through posterior mean estimates  obtained from the above Bayesian inference procedure. We especially wish to check whether the spatiotemporal field of first arrival dates and the niche of the species are identifiable. This verification is important for the proposed models owing to their high structural complexity with a relatively large number of parameters. For the simulation study, we mimic our data application and choose the same spatial domain and temporal replicates, i.e., we simulate datasets at the $D=1268$ pixels over the $T=21$ years from 2001 to 2021 with the same covariates and data structures, including the climate covariate (NAO), the number of BBS stops and the median duration in each pixel-year. The values we set for latent Gaussian processes and hyperparameters are similar to those obtained when having fitted this model to \REV{the species \textit{Chimney Swift}}. \REV{To test model robustness, we allow for potential model misspecification and overdispersion in the checklist counts. So we add another layer of randomness to the count component, giving 
\begin{align*}
	N^{\text{ckl}}_i \mid \Lambda^{\text{ckl}}_i, \bm \theta_\text{ckl} &\sim \mathrm{Pois} (\Lambda^{\text{ckl}}_i), \\
 \Lambda^{\text{ckl}}_i &\sim \Gamma\{r, r \lambda^{\text{ckl}}(\bm s_i, t_i ; \bm \theta_\text{ckl}) \},
\end{align*} 
where $\Gamma(a,b)$ is the Gamma distribution with shape $a>0$ and scale $b>0$. As a result,  $N^{\text{ckl}}_i$ has negative binomial distribution with mean $\lambda^{\text{ckl}}(\bm s_i, t_i )$ and an inflated variance $\lambda^{\text{ckl}}(\bm s_i, t_i ) + \lambda^{\text{ckl}}(\bm s_i, t_i )^2/r$. We set $r=10$ and estimate all the parameters for $100$ independent replicates of such data.} 




The different model components are generally well identified and reproduced by our estimation procedure.  The results in the Appendix (Figure~\ref{fig:simulation}) suggest that the posterior means of the median first arrival dates estimated with this procedure are indeed essentially unbiased, and the truth is mostly well recovered. Importantly, the temporal evolution and spatial patterns are also captured satisfactorily. For example, in our  simulation setting we  have spatially more homogeneous and later first arrivals for the earlier years such as 2003, and the estimated posterior means reproduce this property. The boxplots in Figure~\ref{fig:simulation} suggest that first arrivals for certain pixels are harder to estimate than others, but the truth falls within the predicted interquartile ranges in most cases. \REV{We obtained similar simulation results in the setting where there is no model misspecification (not shown), i.e., $r=\infty$.}




\section{Results}\label{sec:results}

We now apply the model constructed and checked in Sections~\ref{sec:model} and~\ref{sec:bayesian} to the bird data described in Section~\ref{sec:data}, where we run separate models for the $10$ species with largest occurrence numbers in eBird.   Due to space constraints, we report a selection of results  for four species: \textit{Chimney Swift}, \textit{Great Crested Flycatcher}, \textit{Chestnut-sided Warbler} and \textit{Purple Martin}. The MCMC trace plots in the Appendix (Figure \ref{fig:trace}) for one of the aforementioned species indicate that the chains are mixing relatively well, though some parameters like $\theta^\text{act}$ and $\beta_0^{\text{GEV-}\mu}$ still display autocorrelation after the burn-in period and thinning.

\REV{To compare our approach with a simpler one where only the first arrival dates are considered but without modelling the niche or sampling effort, we also report results for a model using only the $Z_i$-variables but  not $N_i^{\text{BBS}}$, $N_i^{\text{ckl}}$ and $N_i^{\text{spc}}$. We fit this model separately and call it the \emph{GEV-only} model, for which $\theta^{\text{niche-GEV}}$ is null and $x_{\text{effort}}$ set to a constant value.}

\begin{figure*}[t]
    \centering
 \begin{subfigure}{0.85\textwidth}
\includegraphics[width=.99\textwidth]{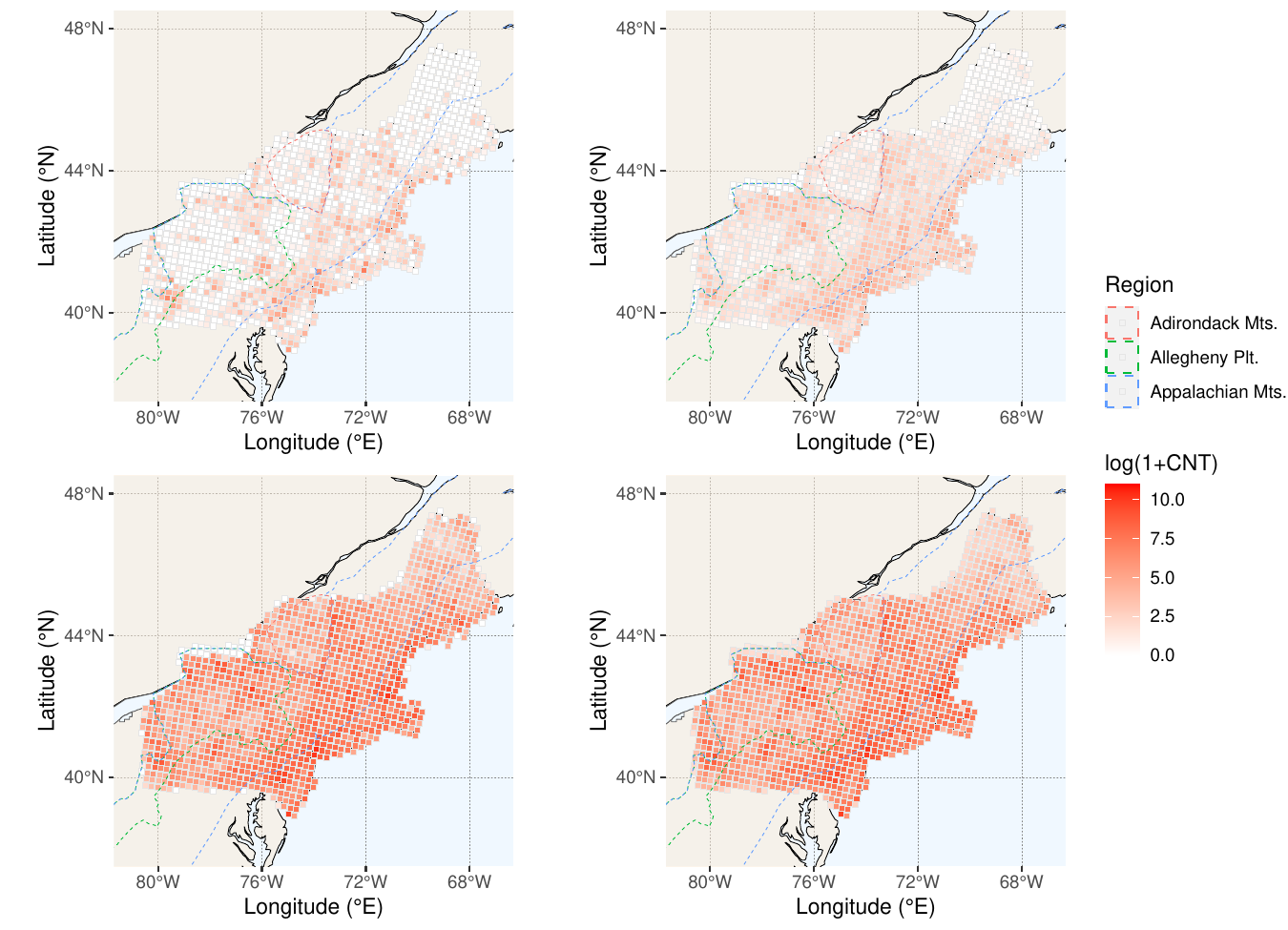}
\end{subfigure} \\
  \caption{Observed numbers of checklists, $N^\text{ckl}_i$ (left) against model-based  posterior mean estimates of the Poisson mean number of checklists, $\lambda^\text{ckl}(\bm s_i, t_i)$ (right), using the model fitted to \textit{Chimney Swift} in 2001 (top) and 2021 (bottom).}
  \label{fig:posterior:cox}
\end{figure*}

Figure \ref{fig:posterior:cox} illustrates that there have been more checklists in recent years, and the model captures these temporal differences and also the  spatial pattern in observed checklist numbers. 

The posterior means of the sharing parameters are all positive for $\theta^\text{pref}$  and \REV{mostly} negative (or positive with $95\%$ posterior credible interval including zero) for $\theta^\text{act}$ (see Table~\ref{tab:casestudy}), which corroborates our exploratory findings from Section~\ref{sec:data} that higher sampling effort leads to earlier observed dates of first arrival.

\begin{table*}[t]
  \caption{Examples of predictions related to first arrival dates for two pixel-years in 2022. The first \REV{five rows} report the species and the posterior mean estimates of selected parameters, with $95\%$ credible intervals in brackets. The next \REV{rows give the observed, predicted (with the GEV-only model, and with the full model), and  debiased (with the full model) median} first arrival dates (in day/month format)  at the two pixels indicated by the two triangles in Figure \ref{fig:posterior:binomial}); the first \REV{four rows here} correspond to the pixel in the southeastern area, the last \REV{four rows} to the pixel located in the middle of the study region. An ``NA'' value in the Observed row indicates that there were no observations within that pixel in 2022. The strikethroughs over the dates indicate that our full model estimates a  species presence probability of less than $1\%$ for that pixel-year.}
    \centering
    \begin{tabular}{l|cccc }
   Species  &  Chimney Swift &  Great Crested Flycatcher & Chestnut-sided Warbler &  Purple Martin \\
    \hline
   $\hat{\theta}^\text{pref}$& 0.191 (0.184,0.202)& 0.204 (0.199,0.21) & 0.187 (0.183,0.191) & 0.2 (0.178,0.217) \\
    $\hat{\theta}^\text{act}$& -0.15 (-0.217,-0.061)& -0.818 (-0.911,-0.696)& -0.548 (-0.619,-0.454) & -0.03 (-0.269,0.236)\\
    $\hat{\theta}^\text{niche-GEV}$  $(\times 10^{-2})$ & 4.9 (4.664,5.134)&4 (3.894,4.133) & 0.2 (0.17,0.278) & 6 (5.541,6.443) \\
    $\beta_1^{\text{GEV-}\mu}$ $(\times 10^{-1})$ & 0.1 (-0.083,0.204)&0.2 (0.125,0.331) & 0.3 (0.295,0.409) & 0 (-0.384,0.526) \\
    \hline
        Observed & NA & NA & NA & NA\\
    Predicted (GEV only) & 28/05 & 27/05 & 01/07 & 18/06\\
        Predicted & 09/05& 03/05& 21/05  & \st{07/06} \\
    Debiased & 03/04& 13/04& 03/05 & \st{28/03} \\
    \hline
   Observed & 01/05  & 04/05  & 04/05  & 29/06 \\
    Predicted (GEV only) & 25/05 & 20/05 & 15/05& 19/05\\
        Predicted & 09/05& 15/05& 12/05  & \st{12/05} \\
    Debiased &  22/04& 05/05& 03/05  & \st{07/04} \\
                 \hline

    \end{tabular}
    \label{tab:casestudy}
\end{table*}

\REV{Table~\ref{tab:casestudy} shows that the sharing parameter $\theta^\text{niche-GEV}$ often has posterior credible intervals that do not contain zero, with most posterior means in the range $[0,0.15]$. This means  that birds tend to be observed earlier in areas where they are relatively abundant.} 
\REV{For $\beta_1^{\text{GEV-}\mu}$, posterior means are generally 
hovering around $[0.01,0.03]$ but often not containing $0$ in its credible intervals, indicating that first arrivals can take place slightly earlier in years with high NAO.}
Higher NAO and temperature anomaly correspond to warmer weather, which could explain earlier arrivals. However, for the majority of the species we investigate these slightly positive NAO coefficients appear not to be significant, as their credible intervals include zero. We conclude that the slight NAO effect on first arrival dates could be present for some species but there is no strong signal of a significant NAO effect across all species; \cite{Wijeyakulasuriya.2023} obtained a similar result when applying their analyses to the \textit{Magnolia Warbler}.   

The GEV shape parameters are negative for all species and have  posterior means around $-0.6$ to $-0.9$, indicating that the estimated first arrival dates have a fixed lower bound.





\subsection{Excursion sets of latent spatial fields}

We study the spatial patterns in the posterior estimation of the four latent spatial fields in \eqref{eq:latent-fields} for the sampling effort, the species niche, and the $\mu$ and $\sigma$ parameters of the distribution of the first arrival dates. 

\begin{figure*}[ht]
    \centering
 \begin{subfigure}{0.85\textwidth}
\includegraphics[width=.99\textwidth]{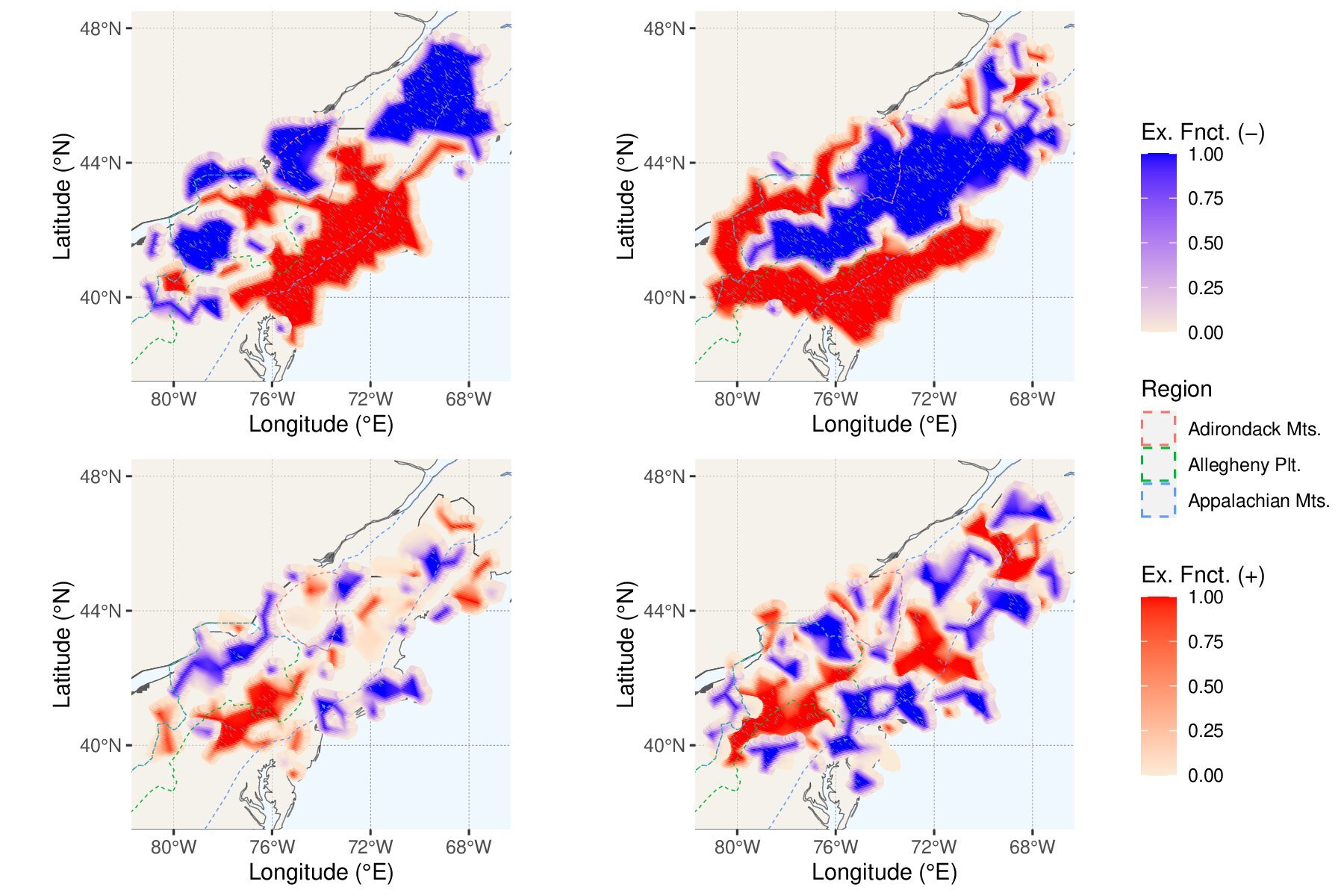}
\end{subfigure} 
  \caption{Excursion set analysis for the four latent spatial fields $X^\text{effort}$ (top left), $X^\text{niche}$ (top right), $X^{\text{GEV-}\mu}$ (bottom left) and $X^{\text{GEV-}\sigma}$ (bottom right) estimated by the model fitted to the \textit{Purple Martin}. Thresholds are fixed at $2$ ,$1$, $0.1$ and $0.1$ for the four fields, using the same order.}
  \label{fig:excursion}
\end{figure*}

To identify areas with very high or low values of the spatial random effects, we utilise credible sets for excursion regions \citep{Bolin-Lindgren.2015}. We evaluate where the fields exceed or fall below certain thresholds. To visualise excursion sets simultaneously for all values of the probability level $\REV{\alpha\in(0,1)}$, \citet{Bolin-Lindgren.2015} introduced the positive and negative excursion functions $F^{+}_{u}(s)  =  1-\inf\{\alpha\mid s\in \mathrm{E}^{+}_{u,\alpha} \}\in [0,1]$ and $F^{-}_{u}(s)  =  1-\inf\{\alpha \mid s\in \mathrm{E}^{-}_{u,\alpha}\} \in[0,1]$\REV{, where $\mathrm{E}^{+}_{u,\alpha}$ is the subregion with maximal surface area in which the spatial field fully exceeds the threshold $u$ with probability $\alpha$ (with an analogous definition of $\mathrm{E}^{-}_{u,\alpha}$ for threshold deficits below $u$)}. 
Figure~\ref{fig:excursion} highlights these excursion functions for the species \emph{Purple Martin}. Sampling effort is high in the southeastern \REV{areas} near the ocean and drops in the further northwestern areas. The niche of the species tends to concentrate in southwestern areas. The fields related to first arrival dates show more fragmented spatial patterns for this species.  


\subsection{Correlation of latent spatial fields with land cover}

To further explain the patterns in the four latent spatial fields, we calculate  the correlation of their posterior means with the land-cover maps described in Section~\ref{sec:data} and shown in the Appendix (Figure~\ref{fig:land-cover}). This can improve our understanding of the land-cover types that drive the four latent processes for a given species. 

\begin{figure}[t]
    \centering
 \begin{subfigure}{0.27\textwidth}
\includegraphics[width=.99\textwidth]{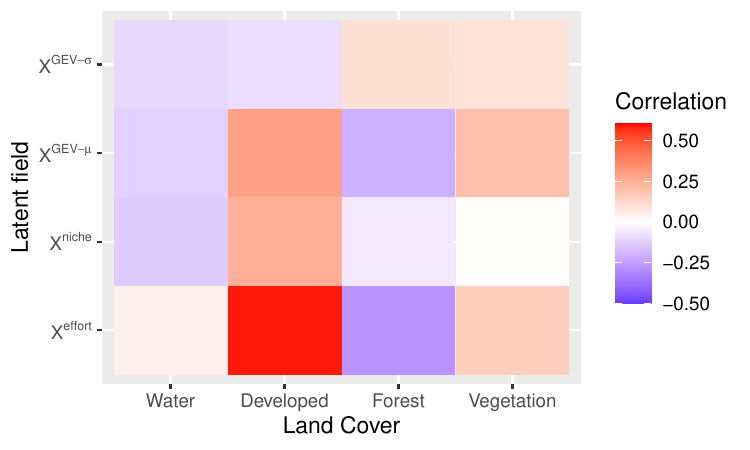}
\end{subfigure} 
    \hspace{-1.3cm}
\begin{subfigure}{0.27\textwidth}
\includegraphics[width=.99\textwidth]{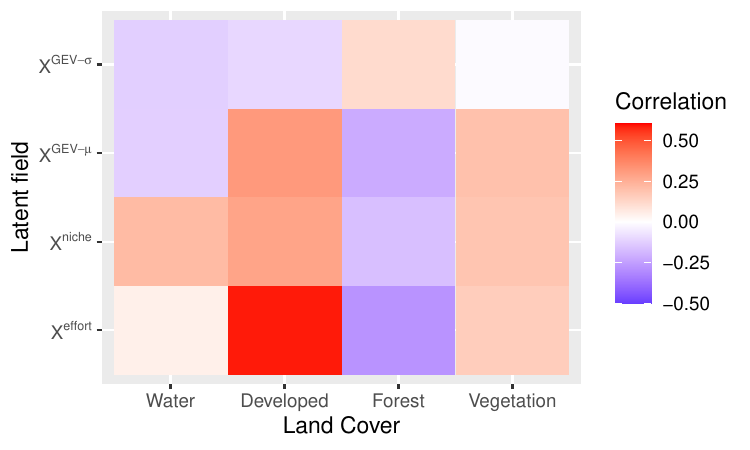}
\end{subfigure} \\
  \begin{subfigure}{0.27\textwidth}
\includegraphics[width=.99\textwidth]{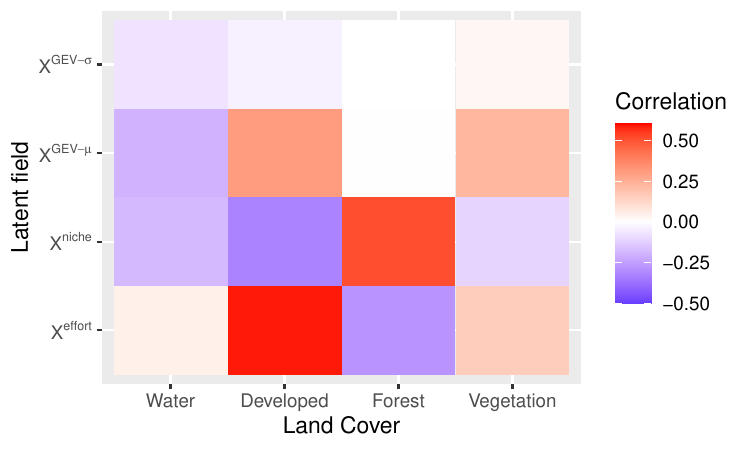}
\end{subfigure} 
    \hspace{-1.3cm}
\begin{subfigure}{0.27\textwidth}
\includegraphics[width=.99\textwidth]{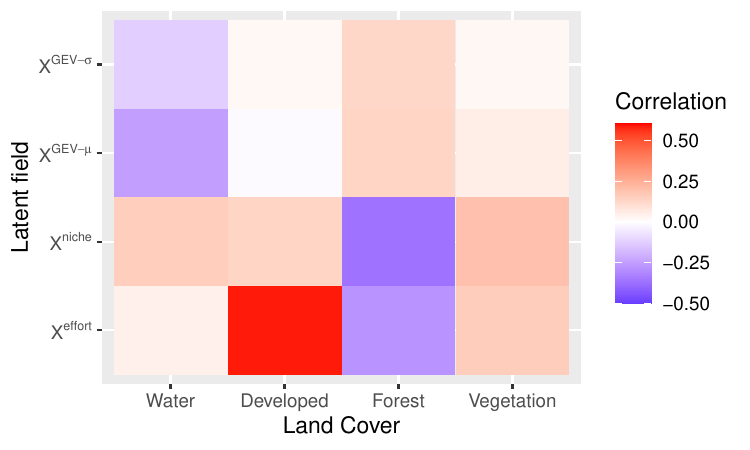}
\end{subfigure} \\
  \caption{Correlations of land-cover proportions (x-axis) with posterior means of the latent fields $X^\text{effort}$, $X^\text{niche}$, $X^{\text{GEV-}\mu}$ and $X^{\text{GEV-}\sigma}$ (y-axis), for four species: \textit{Chimney Swift} (top left), \textit{Great Crested Flycatcher} (top right), \textit{Chestnut-sided Warbler} (bottom left) and \textit{Purple Martin} (bottom right).
  }
  \label{fig:correlation-with-lc}
\end{figure}

Figure~\ref{fig:correlation-with-lc} shows these correlations for four species. The correlations between the estimated latent field for sampling effort and land-cover proportions are very similar across species, which is expected since the checklist count data for the corresponding regression equation are the same in the four models. Sampling effort has strong positive correlation with Developed (built) areas, weaker correlation with Water-dominated or Vegetation-dominated areas, and negative correlation with Forest. Regarding the latent field for niche, it shows different patterns for different species, for example with positive correlations with Forest for species that are known to breed in forests (e.g., \emph{Chestnut-sided Warbler}), and with Developed land cover for species known to breed in Developed areas (e.g., \emph{Chimney Swift}). 
The latent fields describing the spatial variability in first arrival dates show generally weaker correlations with specific land-cover types. For example,  \emph{Chimney Swift} tends to settle earlier in Developed and Vegetation areas than in Water- and Forest-dominated areas.

\subsection{Predicting true first arrivals}\label{sec:case}

\begin{figure*}[t]
    \centering
    \vspace{-0.5cm}
 \begin{subfigure}{0.95\textwidth}
\includegraphics[width=.99\textwidth]{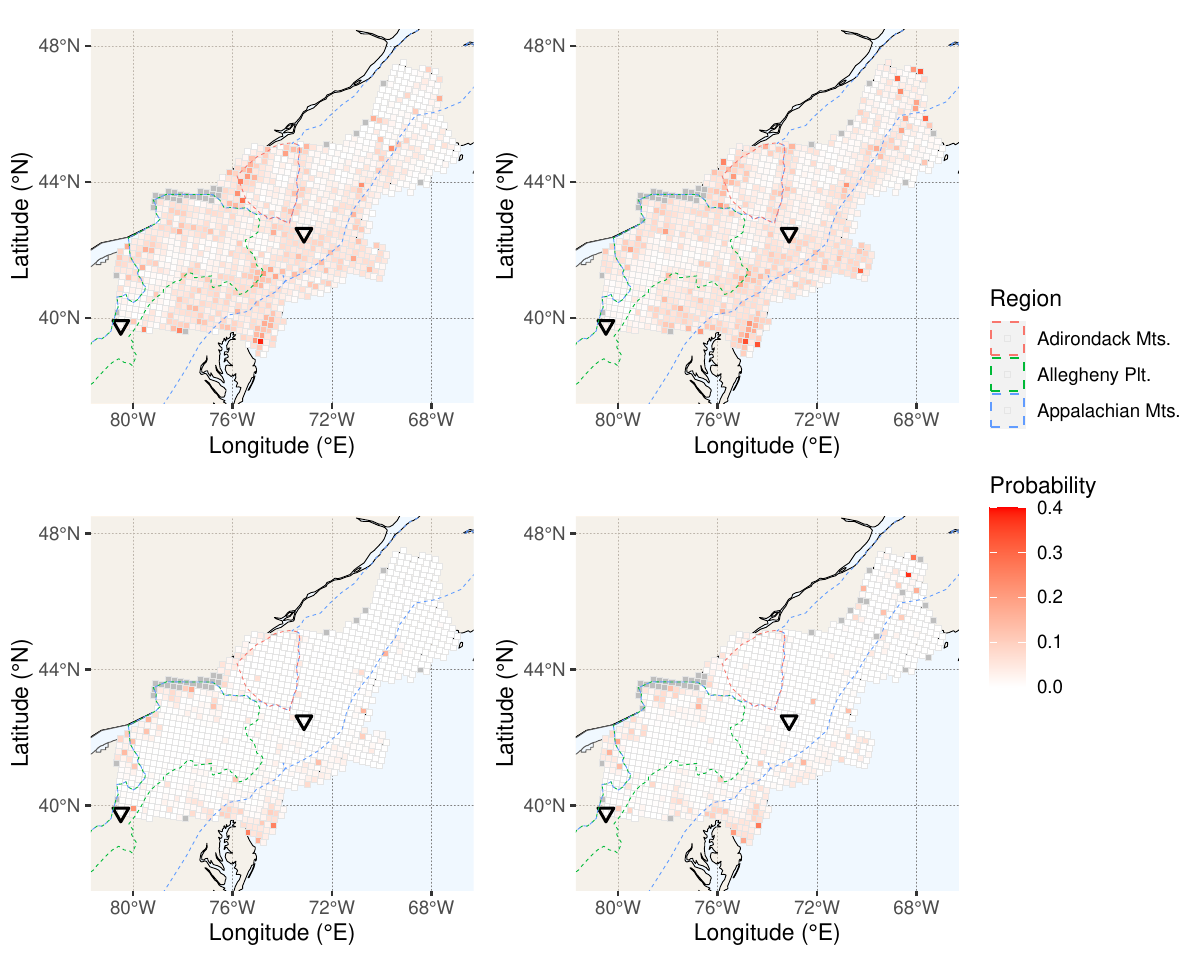}
\end{subfigure} 
  \vspace{-0.5cm}
  \caption{Empirical mean  probability (left displays)  of observing the species \textit{Great Crested Flycatcher} (top) and \textit{Purple Martin} (bottom) in checklists aggregated over all years;  posterior mean probability (right displays) of observing the species from the presence model by using the mean checklist duration aggregated over all years as covariate. The two triangles are the two pixels used in the predictions of first arrival dates detailed in Section~\ref{sec:case}.}
  \label{fig:posterior:binomial}
\end{figure*}

To illustrate the utility of our model, we  predict the true first arrival dates in all pixels. Firstly, Figure \ref{fig:posterior:binomial} shows that the model reproduces generally well the spatial variability in the probability of observing species presence in eBird, with some differences in empirical and model-predicted probabilities due to the additional information provided by BBS data. We use this component of the model to avoid predicting first arrival dates at locations where the presence of the species is uncertain in terms of the estimated niche. Therefore, we do not map predictions of first arrival dates at pixel-years where the estimated presence probability $p^{\text{spc}}\{(\bm s_i,t_i)\}$ is smaller than $1\%$. This feature of our model avoids making predictions in areas where data give no clear signal of a species' breeding activity.

\begin{figure*}[t]
    \centering
 \begin{subfigure}{0.9\textwidth}
\includegraphics[width=.99\textwidth]{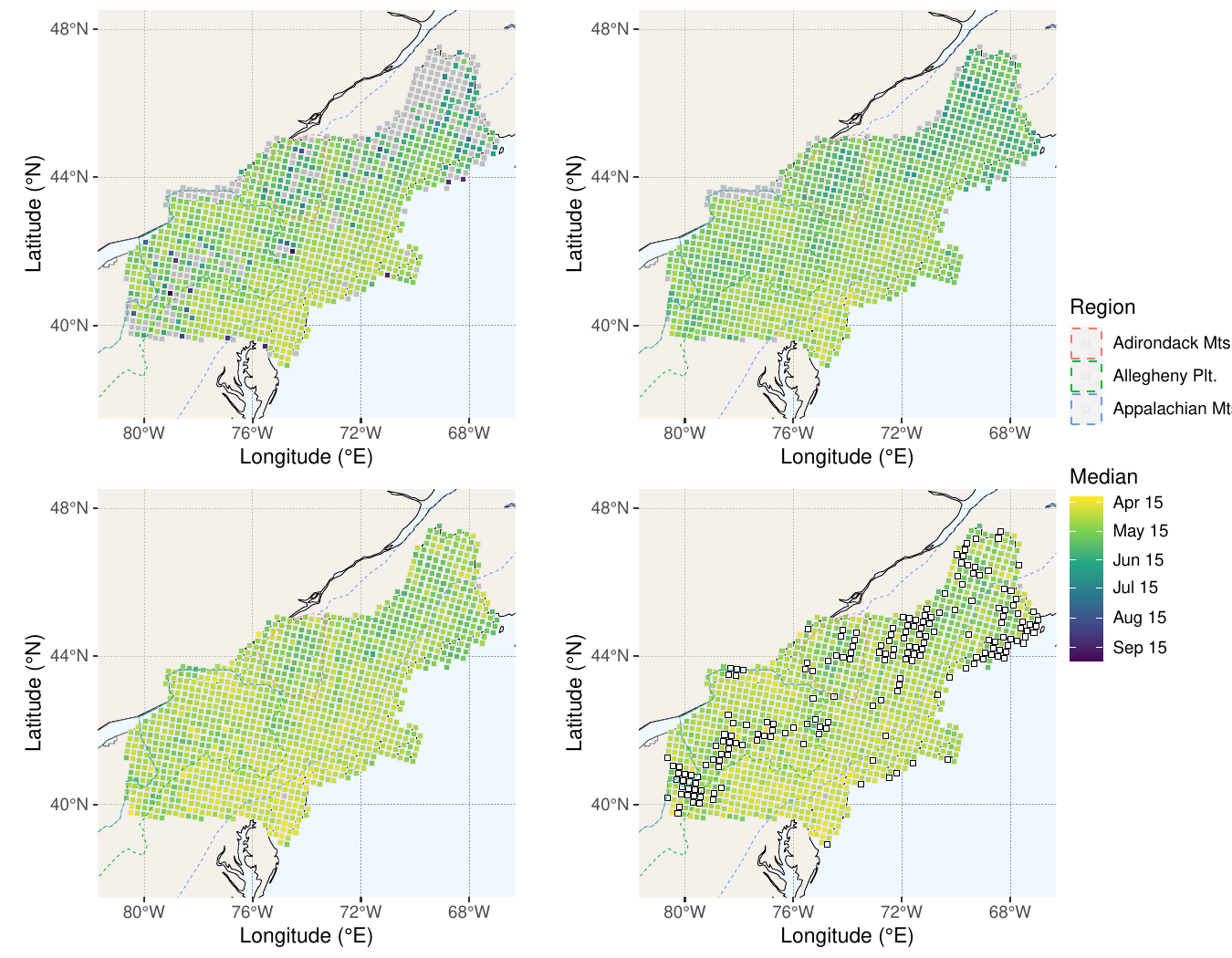}
\end{subfigure} \\
  \caption{ Spatial plots for the species \textit{Great Crested Flycatcher}: Observed first arrivals of \textit{Great Crested Flycatcher} in 2022 (top left), posterior mean of the GEV median first arrivals in 2022 (top right), bias-corrected mean of the GEV median first arrivals in 2022 (bottom left), and the same plot as before though with indications (in white rectangles over the corresponding pixels) for where the model estimates a presence probability of less than $1\%$ for the species (bottom right). The grey pixels in the top left display indicate where no species occurrences were recorded in 2022, or where no checklists and therefore  no median durations were available in that pixel.
  }
  \label{fig:posterior_bias}
\end{figure*}

For the prediction of the true first arrival dates, we set $\lambda^{\text{ckl}}(\bm s_i,t_i)$ and $d_{\bm s_i,t_i}$ to be close to infinity to mimic infinite sampling effort, so we expect to retrieve the true and unbiased dates of first arrival. This approach is illustrated in Figure~\ref{fig:posterior_bias} for the \textit{Great Crested Flycatcher}, where we also show that the model captures the spatial pattern in the observed first arrival dates. 

To expand on our prediction procedure, we focus on two pixels, marked with triangles in Figure~\ref{fig:posterior:binomial}. The first pixel is chosen near the southwestern boundary of our study region and has $83\%$ Forest cover but only $6\%$ Developed land cover. It is located in a rather secluded region with generally low sampling effort. In 2022, there were no reported sightings of the four species of interest in this pixel. Table~\ref{tab:casestudy} highlights that the model can extrapolate 
and draw posterior predictive samples for that pixel-year. The second pixel is chosen in the central area of the study region. For example, a first arrival was observed in 2022 for the \textit{Purple Martin} on the $29^\text{th}$ of June, but Table~\ref{tab:casestudy} (\REV{last two rows}) shows that our model flags this pixel as being uncertain to be located in the  species' niche; \REV{it} is thus possible that in general no breeding activity takes place in this pixel. 
This illustrates the capability of our model to prevent inaccurate extrapolation to spatial regions with low species presence probability. \REV{Table~\ref{tab:casestudy} also shows that the GEV-only model generally estimates a later arrival date than that from the full model after correcting biases. The former predictions usually fall outside of what is expected of the four species' phenology in our study region, i.e., late spring/early summer rather than early spring.} 


\section{Discussion and outlook}\label{sec:conclusion}


\subsection{Statistical learning from participatory data}

Participatory data collection and citizen-science programs can contribute to advancing and sharing scientific knowledge, thanks particularly to the large volume of data and their diversity in terms of spatiotemporal coverage. However, sampling protocols from these sources are usually lenient, with different levels of observer expertise, so statistical techniques are required to correct for heterogeneous sampling effort and observation biases.

In this work, we developed a Bayesian hierarchical model where sampling effort and its influence on the observation bias of a phenological event can be quantified and then corrected during prediction. Our results show that the necessary corrections are often substantial. By singling out the sampling effort, we were also able to estimate spatial probabilities of species presence, which allowed for filtering out implausible regions when providing model-based predictions of the phenological event, since it might never occur. 

\subsection{Bayesian statistics and machine learning}

We carefully constructed statistical models allowing for the inference and interpretation of latent components. Bayesian hierarchical models (BHMs) have many advantages: they permit to identify various latent processes (such as sampling effort) and their interactions, to keep track of uncertainties and assess them through posterior sampling, to incorporate expert knowledge, and to properly separate the process describing how data are observed from the latent ecological processes we seek to infer. 
The Bayesian framework is also relatively robust to noisy data and overdispersion in count data, through the inclusion of appropriately designed random effects. 

Regarding alternative approaches, various machine learning and deep learning tools have found important use in ecology for their skill in predicting observed data, though they may be less effective at capturing complex latent processes.
Further developments are needed to allow such algorithms to predict the processes of interest if these are not directly observed, to extrapolate beyond the range of observed data (as we have done by setting the observation effort to a very high level in predictions), and to avoid propagating or even reinforcing biases from the input data when generating predictive outputs \citep{Dunson2018}.
An interesting goal involves devising mechanisms similar to the sharing of random effects in BHMs but implemented within the architectures of general machine learning and deep learning approaches.

\subsection{Modelling sampling effort}

The eBird sampling effort has various dimensions. We have here considered the annual number of visits to an area (using checklists) and the median duration of  those visits. Understanding how the combination of these two aspects influences the number of detected species individuals is a complex task \citep[e.g., see][]{tang.2021}, and would merit exploration beyond the additive structure of $x_{\text{effort}}$ that we posed in \eqref{eq:x-effort}. Other relevant attributes that we did not consider here are partially reported for eBird checklists, such as the surface of the area where observations took place, or the distance over which the observers moved during the checklist event. Such information could be further studied for a more complete characterisation of the sampling effort and its impact on observation biases with respect to species distributions.  \REV{Identifying sampling effort through Bayesian hierarchical models could also be of interest in various other domains, for example for reports of hail events and damages, and more generally for citizen-science data in climate science.}

\subsection{Detectability of species}

Observation biases related to the detectability of species by the observer remain a major challenge since they are difficult to identify with available data, i.e., it is possible that a present species was not reported. This risk of non detection can vary with land cover, species behaviour, observer experience and other factors. In our model, detectability can be viewed as  being included in the niche process (which could more precisely be called the \emph{detectable niche}), but it is difficult to devise a general statistical approach to further disentangle the niche and the detectability. \REV{Expert knowledge about the conditions of detectability could provide further insights and improve models, but including it was beyond the scope of this work. It would also require designing models adapted to the specific properties of each species and would hinder the intercomparison of models for different species.}

\subsection{Validation of results}

Validation of modelling results is inherently challenging with data that are obtained through heterogeneous sampling efforts and ridden with unknown biases. Certainly, we can validate whether the model appropriately reproduces data as they have been observed. For example, we performed such checks for observed first arrival dates in Section~\ref{sec:results}. However, the latent processes of interest, as well as sampling effort and observation biases, can be entangled in complex ways. Direct validation of inferences on the latent processes of interest is thus difficult, so careful construction of models and inference algorithms is paramount. 

External data sources not suffering from these drawbacks can be used, but they often have much smaller space, time and species coverages. The fusion of data sources, such as in our study with BBS and eBird data, can help improve the estimation of certain model components, but could also be used for validation purposes at space-time locations covered by both datasets. 

Data collection projects are increasingly initiated with advanced sensing technologies that require no human observers, therefore providing new exciting opportunities to collect large amounts of data obeying strict sampling protocols. For instance, one could track bird arrivals with radar data (e.g., \citet{nussbaumer.2021}, or the \url{vogelwarte.ch} project).  Camera traps combined with artificial intelligence for species identification can provide near-continuous temporal coverage, although with rather limited spatial coverage. 



\subsection{Possible extensions of our approach}

Here we focused on identifying past and present spatiotemporal trends in ecological processes. In species distribution modelling, important research  concerns the potential impact of climate change and its interplay with other global change processes (e.g., land use). Our modelling approach could be further refined to establish the relationships between ecological dynamics and climate variables in more detail, and could then be used for projecting first arrival dates under future climate simulations, similar to \citet{Wijeyakulasuriya.2023}.

We have fixed the spatial pixel mesh using a size similar to that routinely used in  modelling results published by the eBird project. It would be interesting to further explore the influence of the mesh size on the properties of the observed first arrival dates. For example, a larger mesh size leads to larger numbers of observations within each pixel, and this potentially reduces the observation bias or better facilitates its quantification. On the other hand, it would also decrease the spatial resolution of predictive maps and \REV{hamper} the identification of relationships of ecological processes with land cover. Further research could 
investigate the role of the mesh choice, or how to optimally combine results from different mesh sizes.

\REV{Regarding temporal trends, the seasonal variation of sampling effort and of the species presence probability, along with how these properties interact with first arrival dates, could 
also be modelled more precisely.}

\REV{If some spatiotemporal domains are visited at high frequency, the corresponding sampling effort could be regarded as being exhaustive, and the structure of the model (e.g., the $g$-function in \eqref{eq:g_function}) could be modified accordingly to improve the identifiability of model components.}

Another extension consists of studying interactions among different species in terms of their niche and phenological events, e.g., to infer which groups of species show similar spring migration  patterns. \REV{This could involve implementing joint species distribution models with each having a set of regression equations and with the possibility to share certain spatial random effects between several species; this could be feasible with our current approach for a small number of species. Alternatively, one could fit models separately for each species as done here, and then apply classification approaches on the posterior estimates of latent model components to group species that show similar model behaviour.} 

\subsection{Ecological extreme-value analysis}

We believe that Extreme-Value Theory could play a significantly more prominent  in ecological science. An important domain of application is the analysis of extreme climatic and environmental events, which are known to have major impacts on biodiversity. A second domain of application, of which the present work is part, concerns the analysis of extreme phenological events using Extreme-Value Theory. Though analyses of this type are still in their early infancy we hope that our work will motivate more widespread applications of extreme-value techniques here.

\section{Acknowledgements}
The authors thank the Editor and anonymous referees for their valuable comments and suggestions. 

\section{Competing interests}
No competing interest is declared.



\section{Data availability }
The data in this article are shared publicly on \url{https://github.com/kohrrelation/mcmc_birds}

\section{Funding}
No funding sources are declared. 


\invisiblesection{s}

\bibliographystyle{chicago}
\bibliography{reference}

\begin{thebibliography}{}

\bibitem[\protect\citeauthoryear{Banerjee, Carlin, and Gelfand}{Banerjee et~al.}{2003}]{Banerjee2003}
Banerjee, S., B.~P. Carlin, and A.~E. Gelfand (2003).
\newblock {\em Hierarchical modeling and analysis for spatial data}.
\newblock Chapman and Hall/CRC.

\bibitem[\protect\citeauthoryear{Bock and Root}{Bock and Root}{1981}]{Bock.1981}
Bock, C.~E. and T.~L. Root (1981).
\newblock The christmas bird count and avian ecology.
\newblock {\em Studies in Avian Biology\/}~{\em 6}, 17--23.

\bibitem[\protect\citeauthoryear{Bolin and Lindgren}{Bolin and Lindgren}{2015}]{Bolin-Lindgren.2015}
Bolin, D. and F.~Lindgren (2015).
\newblock Excursion and contour uncertainty regions for latent {G}aussian models.
\newblock {\em Journal of the Royal Statistical Society: Series B (Statistical Methodology)\/}~{\em 77\/}(1), 85--106.

\bibitem[\protect\citeauthoryear{Bonney, Cooper, and Ballard}{Bonney et~al.}{2016}]{Bonney-2016}
Bonney, R., C.~Cooper, and H.~Ballard (2016, May).
\newblock The theory and practice of citizen science: Launching a new journal.
\newblock {\em Citizen Science: Theory and Practice\/}.

\bibitem[\protect\citeauthoryear{Bonney, Phillips, Ballard, and Enck}{Bonney et~al.}{2016}]{Bonney2016CanCS}
Bonney, R., T.~B. Phillips, H.~L. Ballard, and J.~W. Enck (2016).
\newblock Can citizen science enhance public understanding of science?
\newblock {\em Public Understanding of Science\/}~{\em 25}, 2--16.

\bibitem[\protect\citeauthoryear{Borda-de Água, Alirezazadeh, Neves, Hubbell, Borges, Cardoso, Dionísio, and Pereira}{Borda-de Água et~al.}{2021}]{borda-de-agua.2021}
Borda-de Água, L., S.~Alirezazadeh, M.~Neves, S.~P. Hubbell, P.~A.~V. Borges, P.~Cardoso, F.~Dionísio, and H.~M. Pereira (2021).
\newblock {\em Species Accumulation Curves and Extreme Value Theory}, pp.\  211--226.
\newblock Ecology, Biodiversity and Conservation. Cambridge University Press.

\bibitem[\protect\citeauthoryear{Browne}{Browne}{1996}]{Browne1996.darwin}
Browne, J. (1996).
\newblock Charles darwin: A biography, vol. i, voyaging.
\newblock {\em Journal of the History of Biology\/}~{\em 29\/}(2), 314--316.

\bibitem[\protect\citeauthoryear{Coles}{Coles}{2001}]{Coles.2001}
Coles, S. (2001).
\newblock {\em An introduction to statistical modeling of extreme values}.
\newblock Springer.

\bibitem[\protect\citeauthoryear{Colwell, Coddington, and Hawksworth}{Colwell et~al.}{1994}]{Colwell.1994}
Colwell, R.~K., J.~A. Coddington, and D.~L. Hawksworth (1994).
\newblock Estimating terrestrial biodiversity through extrapolation.
\newblock {\em Philosophical Transactions of the Royal Society of London. Series B: Biological Sciences\/}~{\em 345\/}(1311), 101--118.

\bibitem[\protect\citeauthoryear{Conklin, Lisovski, and Battley}{Conklin et~al.}{2021}]{Conklin.etal.2021}
Conklin, J.~R., S.~Lisovski, and P.~F. Battley (2021).
\newblock Advancement in long-distance bird migration through individual plasticity in departure.
\newblock {\em Nature Communications\/}~{\em 12\/}(1), 4780.

\bibitem[\protect\citeauthoryear{{Cornell Lab of Ornithology}}{{Cornell Lab of Ornithology}}{2022}]{eBird}
{Cornell Lab of Ornithology} (2022).
\newblock {eBird Basic Dataset, Version: EBD relNov-2022}.

\bibitem[\protect\citeauthoryear{Cotton}{Cotton}{2003}]{Cotton.2003}
Cotton, P.~A. (2003).
\newblock Avian migration phenology and global climate change.
\newblock {\em Proceedings of the National Academy of Sciences\/}~{\em 100\/}(21), 12219--12222.

\bibitem[\protect\citeauthoryear{Datta, Banerjee, Finley, and Gelfand}{Datta et~al.}{2016}]{Datta2016}
Datta, A., S.~Banerjee, A.~O. Finley, and A.~E. Gelfand (2016).
\newblock Hierarchical nearest-neighbor gaussian process models for large geostatistical datasets.
\newblock {\em Journal of the American Statistical Association\/}~{\em 111\/}(514), 800--812.

\bibitem[\protect\citeauthoryear{Diggle, Menezes, and Su}{Diggle et~al.}{2010}]{Diggle2010}
Diggle, P.~J., R.~Menezes, and T.-l. Su (2010).
\newblock Geostatistical inference under preferential sampling.
\newblock {\em Journal of the Royal Statistical Society: Series C (Applied Statistics)\/}~{\em 59\/}(2), 191--232.

\bibitem[\protect\citeauthoryear{Dunson}{Dunson}{2018}]{Dunson2018}
Dunson, D.~B. (2018).
\newblock Statistics in the big data era: Failures of the machine.
\newblock {\em Statistics \& Probability Letters\/}~{\em 136}, 4--9.

\bibitem[\protect\citeauthoryear{Fink, Auer, Johnston, Ruiz-Gutierrez, Hochachka, and Kelling}{Fink et~al.}{2020}]{Fink2020}
Fink, D., T.~Auer, A.~Johnston, V.~Ruiz-Gutierrez, W.~M. Hochachka, and S.~Kelling (2020).
\newblock Modeling avian full annual cycle distribution and population trends with citizen science data.
\newblock {\em Ecological Applications\/}~{\em 30\/}(3), e02056.

\bibitem[\protect\citeauthoryear{Fraisl, Campbell, See, Wehn, Wardlaw, Gold, Moorthy, Arias, Piera, Oliver, Mas{\'o}, Penker, and Fritz}{Fraisl et~al.}{2020}]{Frisl.2020.UN}
Fraisl, D., J.~Campbell, L.~See, U.~Wehn, J.~Wardlaw, M.~Gold, I.~Moorthy, R.~Arias, J.~Piera, J.~L. Oliver, J.~Mas{\'o}, M.~Penker, and S.~Fritz (2020).
\newblock Mapping citizen science contributions to the un sustainable development goals.
\newblock {\em Sustainability Science\/}~{\em 15\/}(6), 1735--1751.

\bibitem[\protect\citeauthoryear{Fraisl, Hager, Bedessem, Gold, Hsing, Danielsen, Hitchcock, Hulbert, Piera, Spiers, Thiel, and Haklay}{Fraisl et~al.}{2022}]{Fraisl.2022}
Fraisl, D., G.~Hager, B.~Bedessem, M.~Gold, P.-Y. Hsing, F.~Danielsen, C.~B. Hitchcock, J.~M. Hulbert, J.~Piera, H.~Spiers, M.~Thiel, and M.~Haklay (2022).
\newblock Citizen science in environmental and ecological sciences.
\newblock {\em Nature Reviews Methods Primers\/}~{\em 2\/}(1), 64.

\bibitem[\protect\citeauthoryear{Fuglstad, Simpson, Lindgren, and Rue}{Fuglstad et~al.}{2018}]{Fuglstad.al.2018}
Fuglstad, G.-A., D.~Simpson, F.~Lindgren, and H.~Rue (2018).
\newblock Constructing priors that penalize the complexity of {G}aussian random fields.
\newblock {\em Journal of the American Statistical Association\/}~{\em 114\/}(525), 445--452.

\bibitem[\protect\citeauthoryear{Gaines and Denny}{Gaines and Denny}{1993}]{Gaines1993}
Gaines, S.~D. and M.~W. Denny (1993).
\newblock The largest, smallest, highest, lowest, longest, and shortest: extremes in ecology.
\newblock {\em Ecology\/}~{\em 74\/}(6), 1677--1692.

\bibitem[\protect\citeauthoryear{Gelfand and Shirota}{Gelfand and Shirota}{2019}]{Gelfand2019}
Gelfand, A.~E. and S.~Shirota (2019).
\newblock Preferential sampling for presence/absence data and for fusion of presence/absence data with presence-only data.
\newblock {\em Ecological Monographs\/}~{\em 89\/}(3), e01372.

\bibitem[\protect\citeauthoryear{Guinness}{Guinness}{2018}]{Guinness.2018}
Guinness, J. (2018).
\newblock Permutation and grouping methods for sharpening gaussian process approximations.
\newblock {\em Technometrics\/}~{\em 60\/}(4), 415--429.
\newblock PMID: 31447491.

\bibitem[\protect\citeauthoryear{Haklay, Fraisl, Greshake~Tzovaras, Hecker, Gold, Hager, Ceccaroni, Kieslinger, Wehn, Woods, Nold, Balázs, Mazzonetto, Ruefenacht, Shanley, Wagenknecht, Motion, Sforzi, Riemenschneider, Dorler, Heigl, Schaefer, Lindner, Weißpflug, Mačiulienė, and Vohland}{Haklay et~al.}{2021}]{Haklay.2021}
Haklay, M., D.~Fraisl, B.~Greshake~Tzovaras, S.~Hecker, M.~Gold, G.~Hager, L.~Ceccaroni, B.~Kieslinger, U.~Wehn, S.~Woods, C.~Nold, B.~Balázs, M.~Mazzonetto, S.~Ruefenacht, L.~A. Shanley, K.~Wagenknecht, A.~Motion, A.~Sforzi, D.~Riemenschneider, D.~Dorler, F.~Heigl, T.~Schaefer, A.~Lindner, M.~Weißpflug, M.~Mačiulienė, and K.~Vohland (2021).
\newblock Contours of citizen science: a vignette study.
\newblock {\em Royal Society Open Science\/}~{\em 8\/}(8), 202108.

\bibitem[\protect\citeauthoryear{Hitz, Davis, and Samorodnitsky}{Hitz et~al.}{2017}]{Hitz2017}
Hitz, A., R.~Davis, and G.~Samorodnitsky (2017).
\newblock Discrete extremes.
\newblock {\em arXiv preprint arXiv:1707.05033\/}.

\bibitem[\protect\citeauthoryear{Hsing, Hill, Smith, Bradley, Green, Kent, Mason, Rees, Whittingham, Cokill, Scientists, and Stephens}{Hsing et~al.}{2022}]{Hsing.eMammal.2022}
Hsing, P.-Y., R.~A. Hill, G.~C. Smith, S.~Bradley, S.~E. Green, V.~T. Kent, S.~S. Mason, J.~Rees, M.~J. Whittingham, J.~Cokill, M.~C. Scientists, and P.~A. Stephens (2022).
\newblock Large-scale mammal monitoring: The potential of a citizen science camera-trapping project in the united kingdom.
\newblock {\em Ecological Solutions and Evidence\/}~{\em 3\/}(4), e12180.

\bibitem[\protect\citeauthoryear{Huser, Stein, and Zhong}{Huser et~al.}{2023}]{Huser.vecchia.2023}
Huser, R., M.~L. Stein, and P.~Zhong (2023).
\newblock Vecchia likelihood approximation for accurate and fast inference with intractable spatial max-stable models.
\newblock {\em Journal of Computational and Graphical Statistics. In press.\/}.

\bibitem[\protect\citeauthoryear{Isaac, van Strien, August, de~Zeeuw, and Roy}{Isaac et~al.}{2014}]{Isaac.2014}
Isaac, N. J.~B., A.~J. van Strien, T.~A. August, M.~P. de~Zeeuw, and D.~B. Roy (2014).
\newblock Statistics for citizen science: extracting signals of change from noisy ecological data.
\newblock {\em Methods in Ecology and Evolution\/}~{\em 5\/}(10), 1052--1060.

\bibitem[\protect\citeauthoryear{Johnston, Fink, Hochachka, and Kelling}{Johnston et~al.}{2018}]{Johnston.2018}
Johnston, A., D.~Fink, W.~M. Hochachka, and S.~Kelling (2018).
\newblock Estimates of observer expertise improve species distributions from citizen science data.
\newblock {\em Methods in Ecology and Evolution\/}~{\em 9\/}(1), 88--97.

\bibitem[\protect\citeauthoryear{Johnston, Hochachka, Strimas-Mackey, Ruiz~Gutierrez, Robinson, Miller, Auer, Kelling, and Fink}{Johnston et~al.}{2021}]{Johnston2021}
Johnston, A., W.~M. Hochachka, M.~E. Strimas-Mackey, V.~Ruiz~Gutierrez, O.~J. Robinson, E.~T. Miller, T.~Auer, S.~T. Kelling, and D.~Fink (2021).
\newblock Analytical guidelines to increase the value of community science data: An example using ebird data to estimate species distributions.
\newblock {\em Diversity and Distributions\/}~{\em 27\/}(7), 1265--1277.

\bibitem[\protect\citeauthoryear{Johnston, Matechou, and Dennis}{Johnston et~al.}{2023}]{Johnston2023}
Johnston, A., E.~Matechou, and E.~B. Dennis (2023).
\newblock Outstanding challenges and future directions for biodiversity monitoring using citizen science data.
\newblock {\em Methods in Ecology and Evolution\/}~{\em 14\/}(1), 103--116.

\bibitem[\protect\citeauthoryear{Jordan, Ballard, and Phillips}{Jordan et~al.}{2012}]{Jordan.2012}
Jordan, R.~C., H.~L. Ballard, and T.~B. Phillips (2012).
\newblock Key issues and new approaches for evaluating citizen-science learning outcomes.
\newblock {\em Frontiers in Ecology and the Environment\/}~{\em 10\/}(6), 307--309.

\bibitem[\protect\citeauthoryear{Katz, Brush, and Parlange}{Katz et~al.}{2005}]{Katz2005}
Katz, R.~W., G.~S. Brush, and M.~B. Parlange (2005).
\newblock Statistics of extremes: modeling ecological disturbances.
\newblock {\em Ecology\/}~{\em 86\/}(5), 1124--1134.

\bibitem[\protect\citeauthoryear{Katzfuss and Guinness}{Katzfuss and Guinness}{2021}]{Katzfuss.Guinness.2021}
Katzfuss, M. and J.~Guinness (2021).
\newblock {A General Framework for Vecchia Approximations of Gaussian Processes}.
\newblock {\em Statistical Science\/}~{\em 36\/}(1), 124 -- 141.

\bibitem[\protect\citeauthoryear{Kelling, Johnston, Hochachka, Iliff, Fink, Gerbracht, Lagoze, La~Sorte, Moore, Wiggins, et~al.}{Kelling et~al.}{2015}]{Kelling2015}
Kelling, S., A.~Johnston, W.~M. Hochachka, M.~Iliff, D.~Fink, J.~Gerbracht, C.~Lagoze, F.~A. La~Sorte, T.~Moore, A.~Wiggins, et~al. (2015).
\newblock Can observation skills of citizen scientists be estimated using species accumulation curves?
\newblock {\em PloS one\/}~{\em 10\/}(10), e0139600.

\bibitem[\protect\citeauthoryear{Koh, Pimont, Dupuy, and Opitz}{Koh et~al.}{2023}]{Koh.2022}
Koh, J., F.~Pimont, J.-L. Dupuy, and T.~Opitz (2023).
\newblock Spatiotemporal wildfire modelling through point processes with moderate and extreme marks.
\newblock {\em The Annals of Applied Statistics\/}~{\em 17\/}(1), 560--582.

\bibitem[\protect\citeauthoryear{Kosmala, Wiggins, Swanson, and Simmons}{Kosmala et~al.}{2016}]{Kosmala.2016}
Kosmala, M., A.~Wiggins, A.~Swanson, and B.~Simmons (2016).
\newblock Assessing data quality in citizen science.
\newblock {\em Frontiers in Ecology and the Environment\/}~{\em 14\/}(10), 551--560.

\bibitem[\protect\citeauthoryear{Linden}{Linden}{2011}]{Linden2011}
Linden, A. (2011).
\newblock Using first arrival dates to infer bird migration phenology.
\newblock {\em Boreal Environment Research\/}~{\em 16}, 49--60.

\bibitem[\protect\citeauthoryear{Lindgren, Rue, and Lindstr{\"o}m}{Lindgren et~al.}{2011}]{Lindgren.al.2011}
Lindgren, F., H.~Rue, and J.~Lindstr{\"o}m (2011).
\newblock An explicit link between {G}aussian fields and {G}aussian {M}arkov random fields: the stochastic partial differential equation approach.
\newblock {\em Journal of the Royal Statistical Society: Series B (Statistical Methodology)\/}~{\em 73\/}(4), 423--498.

\bibitem[\protect\citeauthoryear{Majumder, Reich, and Shaby}{Majumder et~al.}{2024}]{Majumder.2024}
Majumder, R., B.~J. Reich, and B.~A. Shaby (2024).
\newblock {Modeling extremal streamflow using deep learning approximations and a flexible spatial process}.
\newblock {\em The Annals of Applied Statistics\/}~{\em 18\/}(2), 1519 -- 1542.

\bibitem[\protect\citeauthoryear{Marshall, Lintott, and Fletcher}{Marshall et~al.}{2015}]{Marshall.2015}
Marshall, P.~J., C.~J. Lintott, and L.~N. Fletcher (2015).
\newblock Ideas for citizen science in astronomy.
\newblock {\em Annual Review of Astronomy and Astrophysics\/}~{\em 53\/}(1), 247--278.

\bibitem[\protect\citeauthoryear{McKinley, Miller-Rushing, Ballard, Bonney, Brown, Evans, French, Parrish, Phillips, Ryan, Shanley, Shirk, Stepenuck, Weltzin, Wiggins, Boyle, Briggs, Chapin~III, Hewitt, Preuss, and Soukup}{McKinley et~al.}{2015}]{McKinley.2015}
McKinley, D.~C., A.~J. Miller-Rushing, H.~L. Ballard, R.~Bonney, H.~Brown, D.~M. Evans, R.~A. French, J.~K. Parrish, T.~B. Phillips, S.~F. Ryan, L.~A. Shanley, J.~L. Shirk, K.~F. Stepenuck, J.~F. Weltzin, A.~Wiggins, O.~D. Boyle, R.~D. Briggs, S.~F. Chapin~III, D.~A. Hewitt, P.~W. Preuss, and M.~A. Soukup (2015).
\newblock Investing in citizen science can improve natural resource management and environmental protection.
\newblock ~{\em 19}.

\bibitem[\protect\citeauthoryear{M{\o}ller, Syversveen, and Waagepetersen}{M{\o}ller et~al.}{1998}]{Moller.al.1998}
M{\o}ller, J., A.~R. Syversveen, and R.~P. Waagepetersen (1998).
\newblock Log gaussian cox processes.
\newblock {\em Scandinavian journal of statistics\/}~{\em 25\/}(3), 451--482.

\bibitem[\protect\citeauthoryear{Newman, Wiggins, Crall, Graham, Newman, and Crowston}{Newman et~al.}{2012}]{Newman.2012}
Newman, G., A.~Wiggins, A.~Crall, E.~Graham, S.~Newman, and K.~Crowston (2012).
\newblock The future of citizen science: emerging technologies and shifting paradigms.
\newblock {\em Frontiers in Ecology and the Environment\/}~{\em 10\/}(6), 298--304.

\bibitem[\protect\citeauthoryear{Noyelle, Robin, Naveau, Yiou, and Faranda}{Noyelle et~al.}{2024}]{Noyelle2024}
Noyelle, R., Y.~Robin, P.~Naveau, P.~Yiou, and D.~Faranda (2024).
\newblock Integration of physical bound constraints to alleviate shortcomings of statistical models for extreme temperatures.
\newblock {\em HAL preprint 04479249\/}.

\bibitem[\protect\citeauthoryear{Nussbaumer, Bauer, Benoit, Mariethoz, Liechti, and Schmid}{Nussbaumer et~al.}{2021}]{nussbaumer.2021}
Nussbaumer, R., S.~Bauer, L.~Benoit, G.~Mariethoz, F.~Liechti, and B.~Schmid (2021).
\newblock Quantifying year-round nocturnal bird migration with a fluid dynamics model.
\newblock {\em Journal of The Royal Society Interface\/}~{\em 18\/}(179), 20210194.

\bibitem[\protect\citeauthoryear{Opitz, Bakka, Huser, and Lombardo}{Opitz et~al.}{2022}]{Opitz2022}
Opitz, T., H.~Bakka, R.~Huser, and L.~Lombardo (2022).
\newblock High-resolution {B}ayesian mapping of landslide hazard with unobserved trigger event.
\newblock {\em The Annals of Applied Statistics\/}~{\em 16\/}(3), 1653--1675.

\bibitem[\protect\citeauthoryear{Pati, Reich, and Dunson}{Pati et~al.}{2011}]{Pati.2011}
Pati, D., B.~J. Reich, and D.~B. Dunson (2011, 03).
\newblock {{B}ayesian geostatistical modelling with informative sampling locations}.
\newblock {\em Biometrika\/}~{\em 98\/}(1), 35--48.

\bibitem[\protect\citeauthoryear{Pocock, Chandler, Bonney, Thornhill, Albin, August, Bachman, Brown, Cunha, Grez, Jackson, Peters, Rabarijaon, Roy, Zaviezo, and Danielsen}{Pocock et~al.}{2018}]{Pocock.2018}
Pocock, M.~J., M.~Chandler, R.~Bonney, I.~Thornhill, A.~Albin, T.~August, S.~Bachman, P.~M. Brown, D.~G.~F. Cunha, A.~Grez, C.~Jackson, M.~Peters, N.~R. Rabarijaon, H.~E. Roy, T.~Zaviezo, and F.~Danielsen (2018).
\newblock Chapter six - a vision for global biodiversity monitoring with citizen science.
\newblock In D.~A. Bohan, A.~J. Dumbrell, G.~Woodward, and M.~Jackson (Eds.), {\em Next Generation Biomonitoring: Part 2}, Volume~59 of {\em Advances in Ecological Research}, pp.\  169--223. Academic Press.

\bibitem[\protect\citeauthoryear{Prieto, G{\'o}mez-D{\'e}niz, and Sarabia}{Prieto et~al.}{2014}]{Prieto2014}
Prieto, F., E.~G{\'o}mez-D{\'e}niz, and J.~M. Sarabia (2014).
\newblock Modelling road accident blackspots data with the discrete generalized {P}areto distribution.
\newblock {\em Accident Analysis \& Prevention\/}~{\em 71}, 38--49.

\bibitem[\protect\citeauthoryear{Ranjbar, Cantoni, Chavez-Demoulin, Marra, Radice, and Jaton}{Ranjbar et~al.}{2022}]{Ranjbar2022}
Ranjbar, S., E.~Cantoni, V.~Chavez-Demoulin, G.~Marra, R.~Radice, and K.~Jaton (2022).
\newblock Modelling the extremes of seasonal viruses and hospital congestion: The example of flu in a {S}wiss hospital.
\newblock {\em Journal of the Royal Statistical Society Series C: Applied Statistics\/}~{\em 71\/}(4), 884--905.

\bibitem[\protect\citeauthoryear{Roberts and Rosenthal}{Roberts and Rosenthal}{2001}]{Roberts.Rosenthal.2001}
Roberts, G.~O. and J.~S. Rosenthal (2001).
\newblock {Optimal scaling for various Metropolis-Hastings algorithms}.
\newblock {\em Statistical Science\/}~{\em 16\/}(4), 351 -- 367.

\bibitem[\protect\citeauthoryear{Rue, Riebler, S{\o}rbye, Illian, Simpson, and Lindgren}{Rue et~al.}{2017}]{Rue.al.2017}
Rue, H., A.~Riebler, S.~H. S{\o}rbye, J.~B. Illian, D.~P. Simpson, and F.~K. Lindgren (2017).
\newblock {Bayesian computing with INLA: a review}.
\newblock {\em Annual Review of Statistics and Its Application\/}~{\em 4}, 395--421.

\bibitem[\protect\citeauthoryear{Silvertown}{Silvertown}{2009}]{SILVERTOWN.2009}
Silvertown, J. (2009).
\newblock A new dawn for citizen science.
\newblock {\em Trends in Ecology \& Evolution\/}~{\em 24\/}(9), 467--471.

\bibitem[\protect\citeauthoryear{Somveille, Rodrigues, and Manica}{Somveille et~al.}{2015}]{Somveille2015}
Somveille, M., A.~S. Rodrigues, and A.~Manica (2015).
\newblock Why do birds migrate? {A} macroecological perspective.
\newblock {\em Global Ecology and Biogeography\/}~{\em 24\/}(6), 664--674.

\bibitem[\protect\citeauthoryear{Tang, Clark, and Gelfand}{Tang et~al.}{2021}]{tang.2021}
Tang, B., J.~S. Clark, and A.~E. Gelfand (2021).
\newblock Modeling spatially biased citizen science effort through the ebird database.
\newblock {\em Environmental and Ecological Statistics\/}~{\em 28\/}(3), 609--630.

\bibitem[\protect\citeauthoryear{Theobald, Ettinger, Burgess, DeBey, Schmidt, Froehlich, Wagner, HilleRisLambers, Tewksbury, Harsch, and Parrish}{Theobald et~al.}{2015}]{THEOBALD.2015}
Theobald, E., A.~Ettinger, H.~Burgess, L.~DeBey, N.~Schmidt, H.~Froehlich, C.~Wagner, J.~HilleRisLambers, J.~Tewksbury, M.~Harsch, and J.~Parrish (2015).
\newblock Global change and local solutions: Tapping the unrealized potential of citizen science for biodiversity research.
\newblock {\em Biological Conservation\/}~{\em 181}, 236--244.

\bibitem[\protect\citeauthoryear{Thibaud, Aalto, Cooley, Davison, and Heikkinen}{Thibaud et~al.}{2016}]{Thibaud2016}
Thibaud, E., J.~Aalto, D.~S. Cooley, A.~C. Davison, and J.~Heikkinen (2016).
\newblock {Bayesian Inference For The Brown-Resnick Process, With An Application To Extreme Low Temperatures}.
\newblock {\em Annals of Applied Statistics\/}~{\em 10\/}(4), 2303--2324.

\bibitem[\protect\citeauthoryear{UN}{UN}{2015}]{UN.2015}
UN (2015).
\newblock A/res/70/1 un general assembly transforming our world: the 2030 agenda for sustainable development.
\newblock {\em Seventieth session of the General Assembly on 25 Sept 2015\/}.

\bibitem[\protect\citeauthoryear{van~de Schoot, Depaoli, King, Kramer, M{\"a}rtens, Tadesse, Vannucci, Gelman, Veen, Willemsen, et~al.}{van~de Schoot et~al.}{2021}]{vandeSchoot2021}
van~de Schoot, R., S.~Depaoli, R.~King, B.~Kramer, K.~M{\"a}rtens, M.~G. Tadesse, M.~Vannucci, A.~Gelman, D.~Veen, J.~Willemsen, et~al. (2021).
\newblock Bayesian statistics and modelling.
\newblock {\em Nature Reviews Methods Primers\/}~{\em 1\/}(1), 1.

\bibitem[\protect\citeauthoryear{Vecchia}{Vecchia}{1988}]{Vecchia.1988}
Vecchia, A.~V. (1988).
\newblock Estimation and model identification for continuous spatial processes.
\newblock {\em Journal of the Royal Statistical Society. Series B (Methodological)\/}~{\em 50\/}(2), 297--312.

\bibitem[\protect\citeauthoryear{Wijeyakulasuriya, Hanks, and Shaby}{Wijeyakulasuriya et~al.}{2023}]{Wijeyakulasuriya.2023}
Wijeyakulasuriya, D.~A., E.~M. Hanks, and B.~A. Shaby (2023).
\newblock Modeling first arrival of migratory birds using a hierarchical max-infinitely divisible process.

\bibitem[\protect\citeauthoryear{Wijeyakulasuriya, Hanks, Shaby, and Cross}{Wijeyakulasuriya et~al.}{2019}]{Wijeyakulasuriya2019}
Wijeyakulasuriya, D.~A., E.~M. Hanks, B.~A. Shaby, and P.~C. Cross (2019).
\newblock Extreme value-based methods for modeling elk yearly movements.
\newblock {\em Journal of Agricultural, Biological and Environmental Statistics\/}~{\em 24\/}(1), 73--91.

\bibitem[\protect\citeauthoryear{Wynn}{Wynn}{2017}]{Wynn_book.2017}
Wynn, J. (2017).
\newblock {\em Citizen science in the digital age : rhetoric, science, and public engagement}.
\newblock Rhetoric, culture, and social critique. Tuscaloosa: The University of Alabama Press Tuscaloosa.

\bibitem[\protect\citeauthoryear{Yadav, Huser, Opitz, and Lombardo}{Yadav et~al.}{2023}]{yadav.2022}
Yadav, R., R.~Huser, T.~Opitz, and L.~Lombardo (2023, 09).
\newblock {Joint modelling of landslide counts and sizes using spatial marked point processes with sub-asymptotic mark distributions}.
\newblock {\em Journal of the Royal Statistical Society Series C: Applied Statistics\/}, qlad077.

\bibitem[\protect\citeauthoryear{Youngflesh, Socolar, Amaral, Arab, Guralnick, Hurlbert, LaFrance, Mayor, Miller, and Tingley}{Youngflesh et~al.}{2021}]{Youngflesh2021}
Youngflesh, C., J.~Socolar, B.~R. Amaral, A.~Arab, R.~P. Guralnick, A.~H. Hurlbert, R.~LaFrance, S.~J. Mayor, D.~A. Miller, and M.~W. Tingley (2021).
\newblock Migratory strategy drives species-level variation in bird sensitivity to vegetation green-up.
\newblock {\em Nature Ecology \& Evolution\/}~{\em 5\/}(7), 987--994.

\end{thebibliography}


\begin{appendices}

\section{Appendix}\label{sec11}

\subsection{Land-cover plots}

\begin{figure*}[!h]
    \centering
    \begin{tabular}{cc}
\includegraphics[width=.38\textwidth]{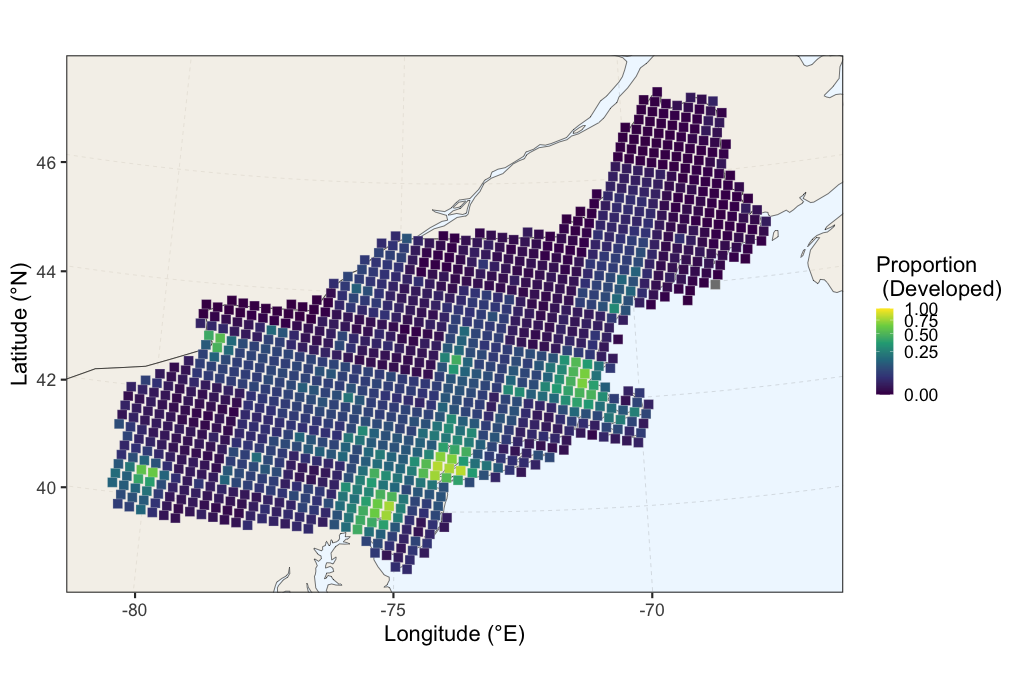} &
\includegraphics[width=.38\textwidth]{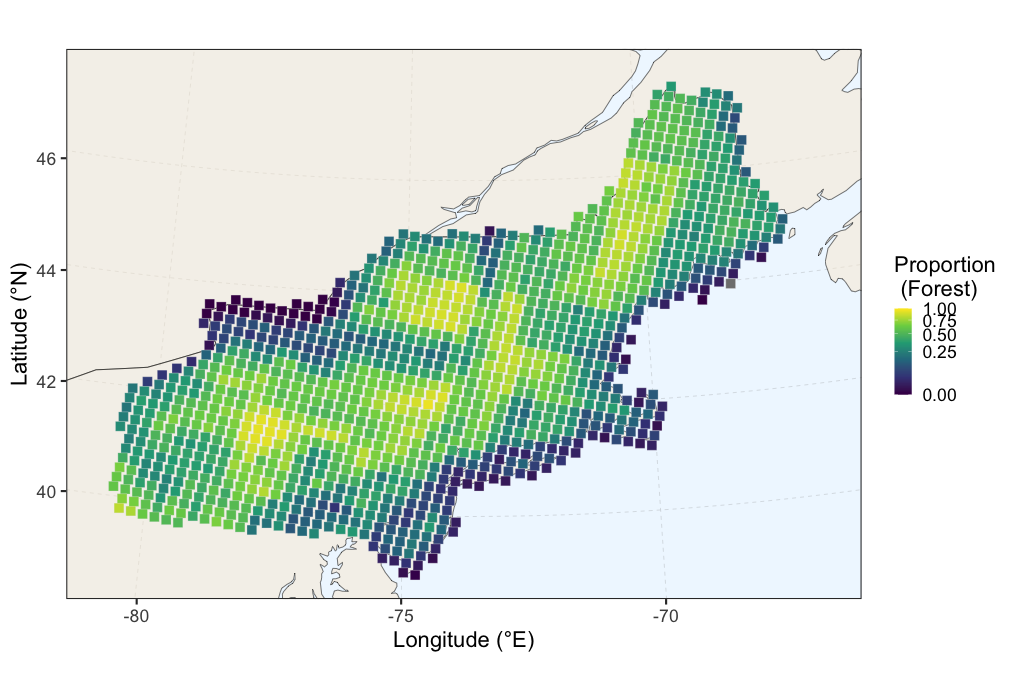}\\
\includegraphics[width=.38\textwidth]{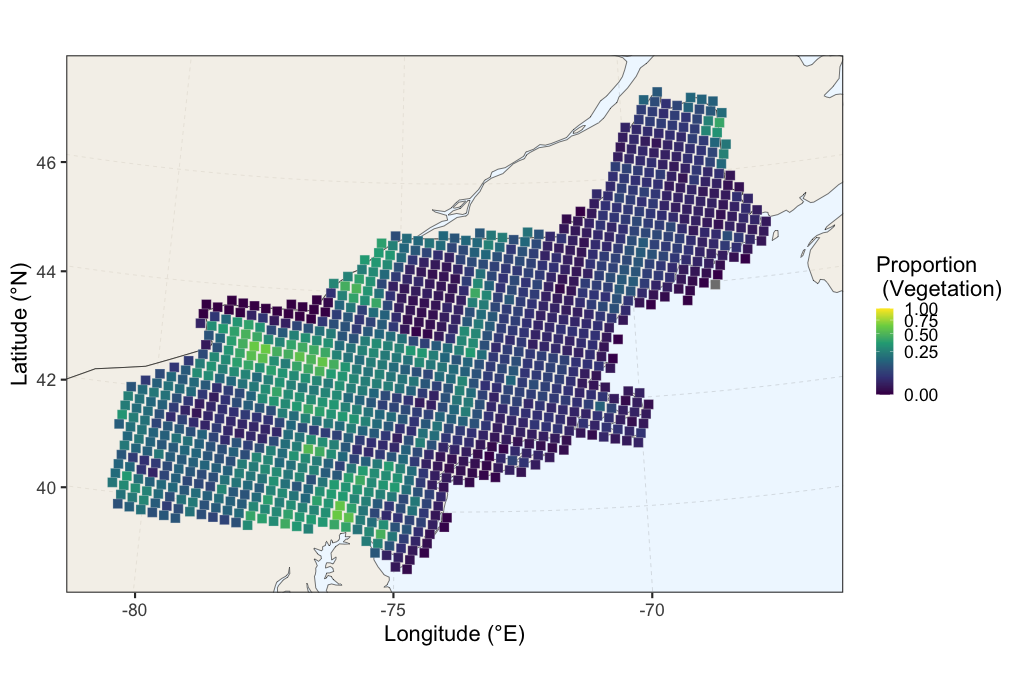} &
\includegraphics[width=.38\textwidth]{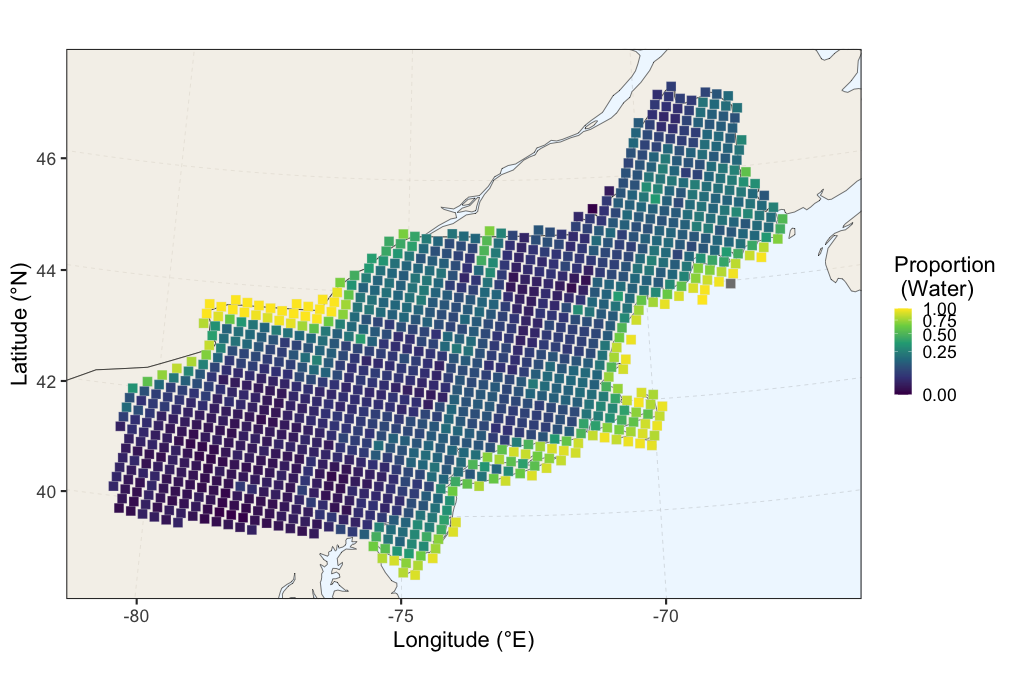}
    \end{tabular}
  \caption{Land-cover proportions in the study area (NLCD, 2021) according to four types described in Section~\ref{sec:data}.}
  \label{fig:land-cover}
\end{figure*}

Figure \ref{fig:land-cover} shows plots of the four land-cover covariates defined as proportions of certain land-cover categories in the pixels of the study area. 

\subsection{Plots from the simulation study}\label{subsec4}

\begin{figure*}[h]
   \centering
\includegraphics[width=.9\textwidth]{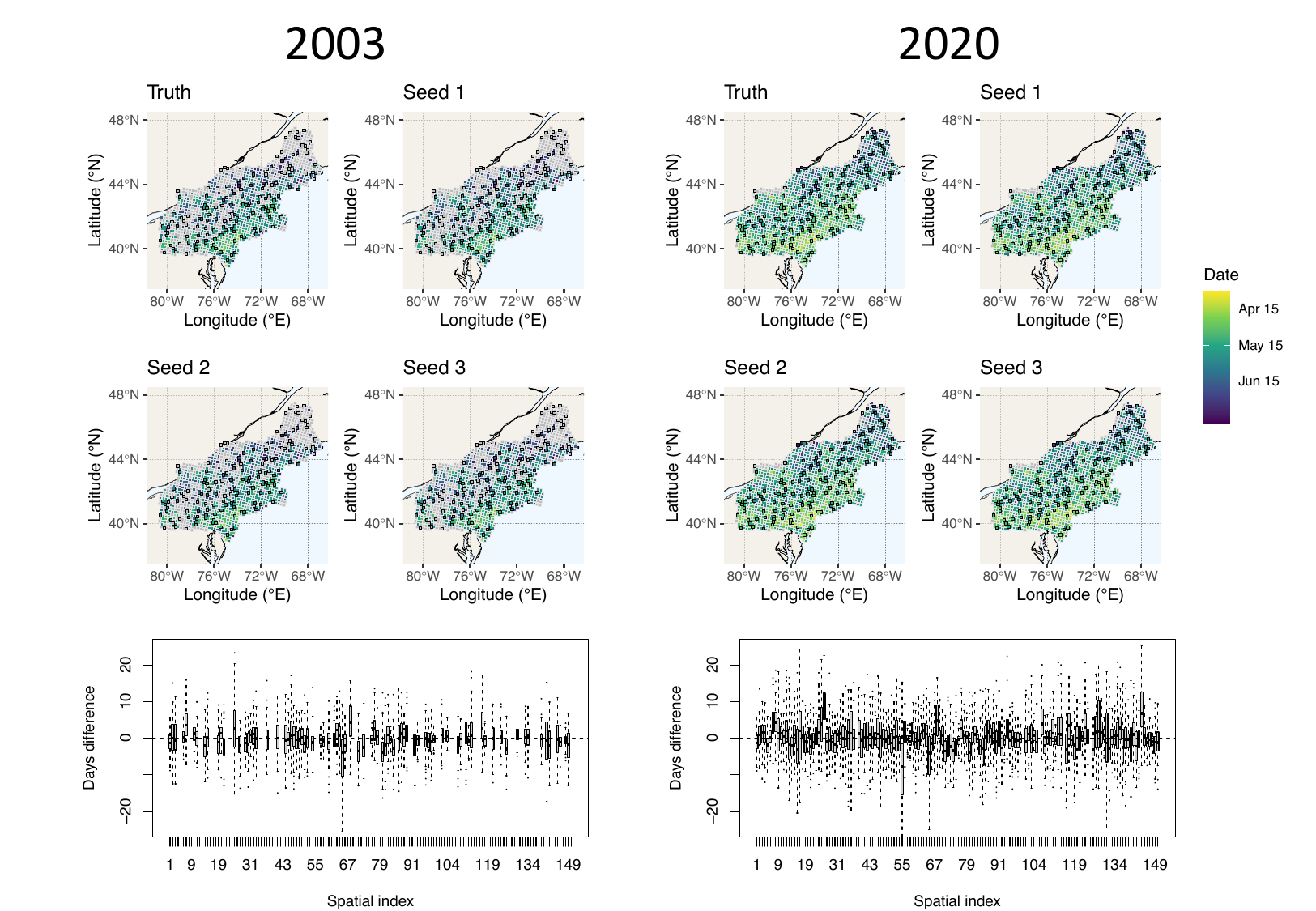}
 \caption{ Simulation study: Truth vs. posterior means for the median first arrival dates in 2003 (left) and 2020 (right), shown with spatial maps (for 3 different seeds in the simulation study), and with boxplots of the differences in days between true and predicted arrival dates, based on 100 simulations. The boxplot shows results for 150 randomly chosen pixels, marked with black boxes in the spatial plots.}
 \label{fig:simulation}
\end{figure*}

Figure \ref{fig:simulation} shows plots assessing the identifiability of the observed first arrival dates in our simulation study.

\subsection{Further plots of  MCMC results}

\begin{figure*}[b]
    \centering
\begin{subfigure}{0.22\textwidth}
\includegraphics[width=.99\textwidth]{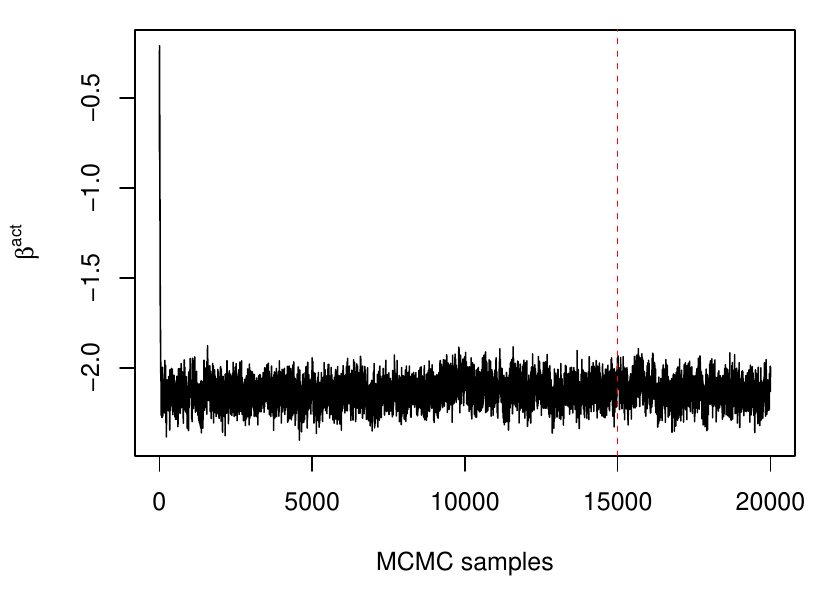}
\end{subfigure}
\begin{subfigure}{0.22\textwidth}
\includegraphics[width=.99\textwidth]{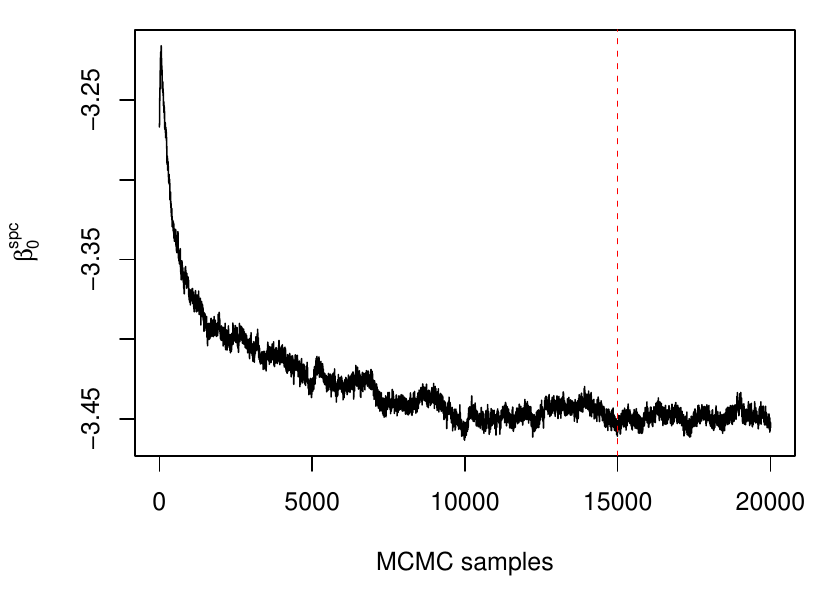}
\end{subfigure} 
\begin{subfigure}{0.22\textwidth}
\includegraphics[width=.99\textwidth]{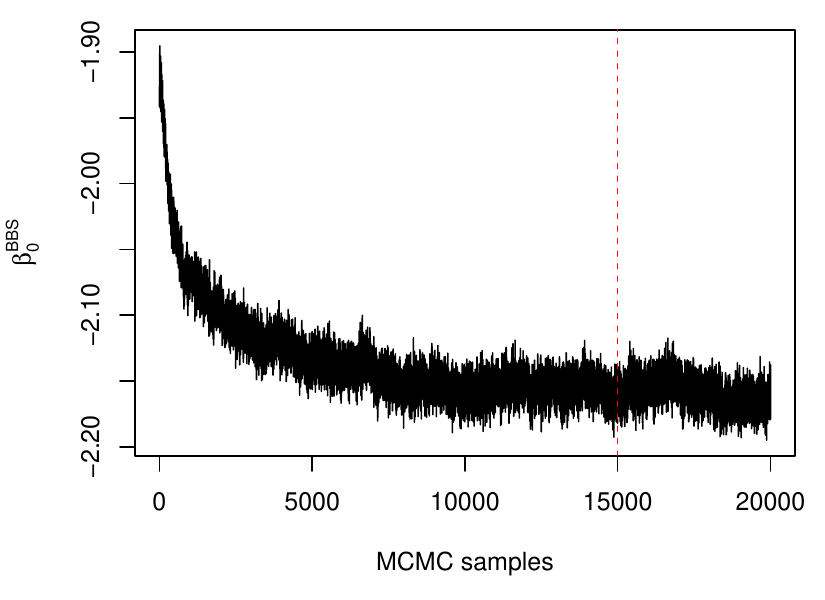}
\end{subfigure}\\
\begin{subfigure}{0.22\textwidth}
\includegraphics[width=.99\textwidth]{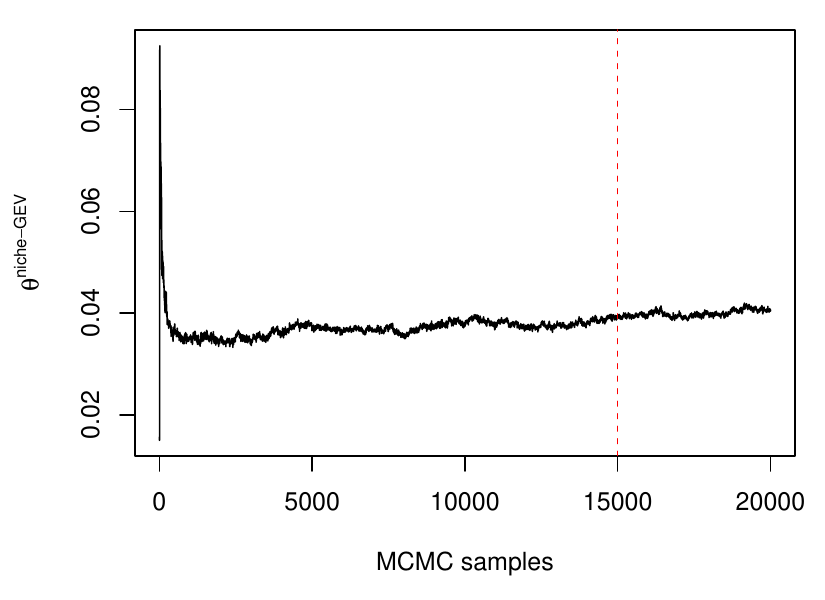}
\end{subfigure}
\begin{subfigure}{0.22\textwidth}
\includegraphics[width=.99\textwidth]{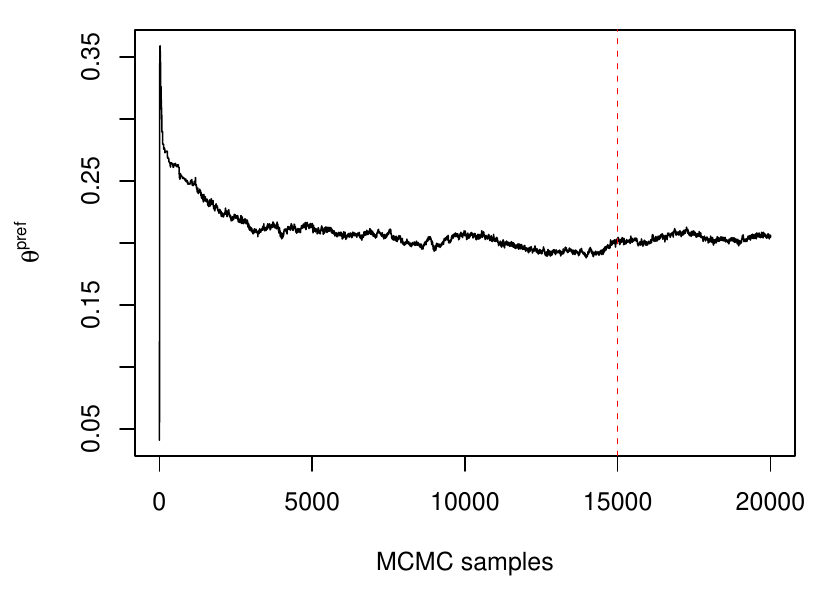}
\end{subfigure} 
\begin{subfigure}{0.22\textwidth}
\includegraphics[width=.99\textwidth]{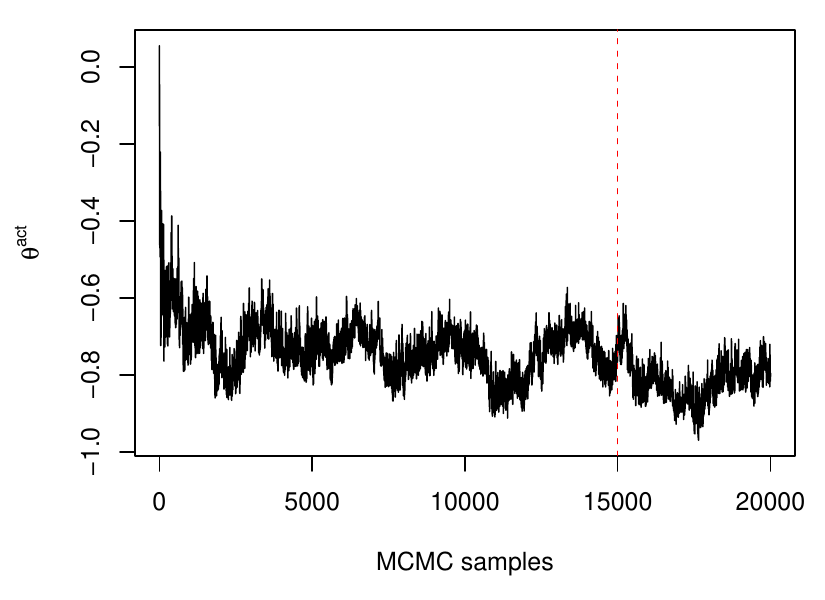}
\end{subfigure}\\
\begin{subfigure}{0.22\textwidth}
\includegraphics[width=.99\textwidth]{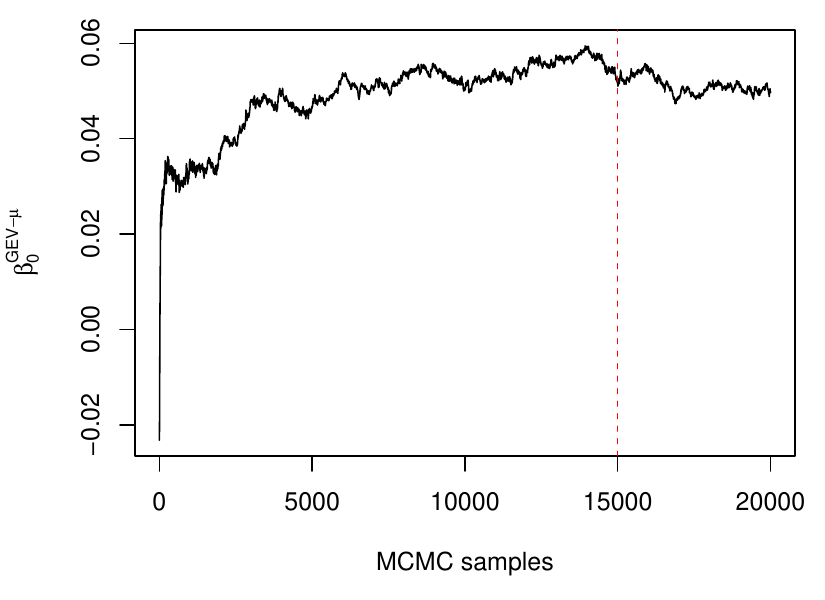}
\end{subfigure} 
\begin{subfigure}{0.22\textwidth}
\includegraphics[width=.99\textwidth]{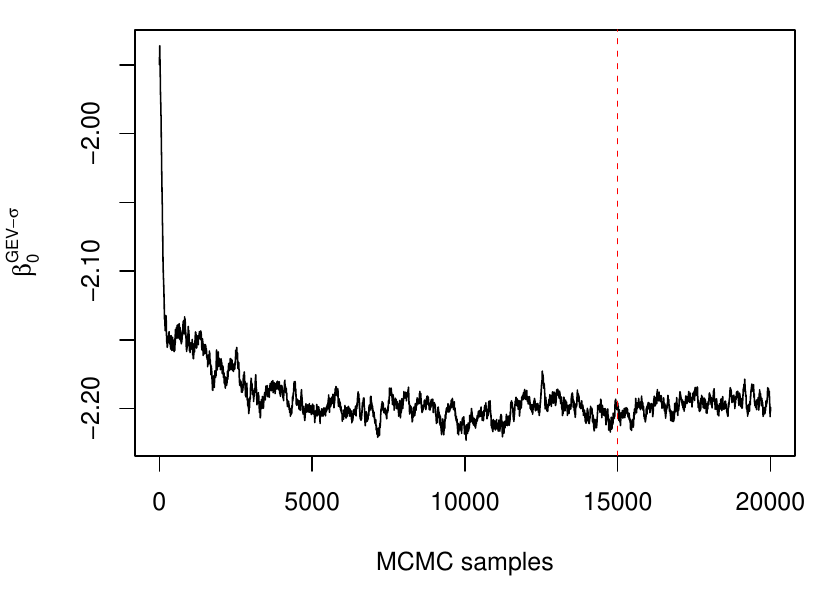}
\end{subfigure}
\begin{subfigure}{0.22\textwidth}
\includegraphics[width=.99\textwidth]{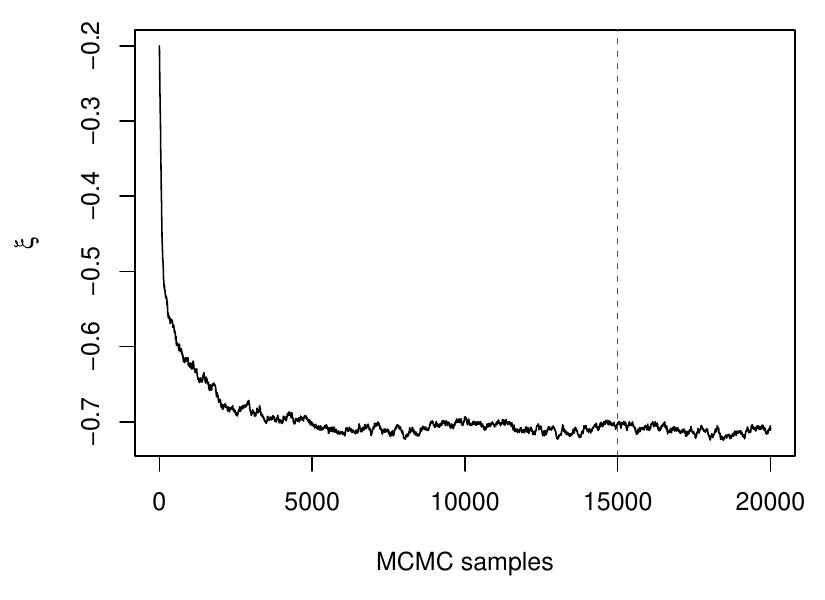}
\end{subfigure}
  \caption{MCMC trace plots for some fixed effect and sharing parameters (top left to bottom right): $\beta^\text{act}$, $\beta_0^\text{spc}$, $\beta_0^\text{BBS}$, $\theta^\text{niche-GEV}$, $\theta^\text{pref}$, $\theta^\text{act}$,  $\beta_0^{\text{GEV-}\mu}$, $\beta_0^{\text{GEV-}\sigma}$ and $\xi$ from our model for the species \textit{Great Crested Flycatcher}. The red dashed line indicates the burn-in period.}
  \label{fig:trace}
\end{figure*}

\begin{figure*}[t]
    \centering
 \begin{subfigure}{0.9\textwidth}
\includegraphics[width=.99\textwidth]{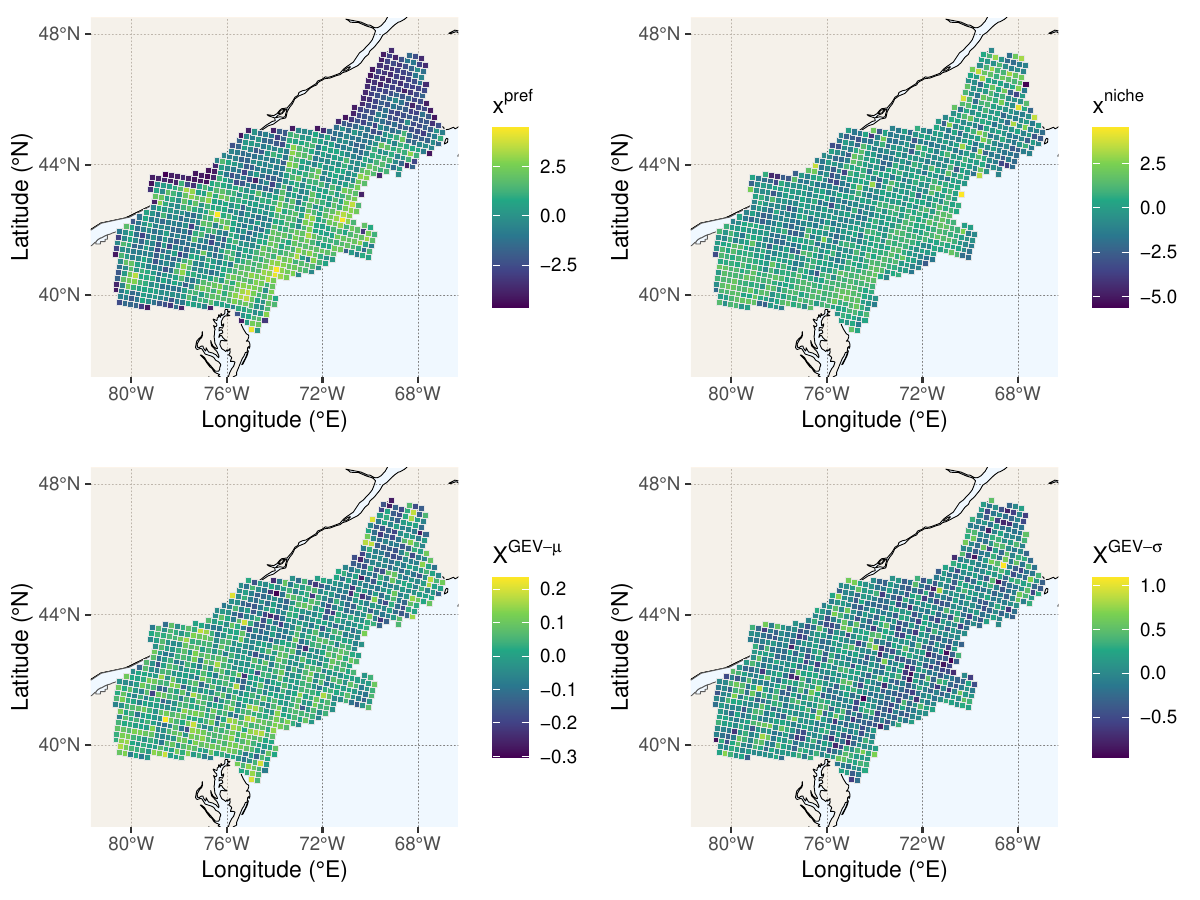}
\end{subfigure} \\
  \caption{Posterior means plots of the four spatial random effects in \eqref{eq:latent-fields} for the model fitted to the species \textit{Chimney Swift}.}
  \label{fig:posterior_spatial}
\end{figure*}

Figure \ref{fig:trace} shows trace plots from the MCMC run of the model on the species \textit{Great Crested Flycatcher}. Figure \ref{fig:posterior_spatial} shows spatial plots of the four latent spatial  fields from the model estimated for the species \textit{Chimney Swift}.

\end{appendices}



\end{document}